\documentclass[reprint, prd]{revtex4-2}
\usepackage{amsmath, amssymb}
\usepackage{graphicx,color,float,tabularx,siunitx}
\usepackage[caption=false]{subfig}
\usepackage{placeins}
\usepackage{rviewport}
\usepackage{orcidlink}
\definecolor{colorLink}{rgb}{0,0,180} 
\usepackage{hyperref}
\hypersetup{
   colorlinks = true,
   citecolor  = colorLink,
   urlcolor   = colorLink,
   linkcolor  = colorLink,
}

\DeclareSIUnit\year{yr}
\usepackage{subfig}
\usepackage{float} % for [H] specifier
\usepackage{booktabs}
\usepackage{multirow}
\usepackage{siunitx}

\usepackage[sort&compress]{natbib}

\usepackage[shortcuts]{extdash}

\begin{document}
\title{Signatures of NED on Quasi periodic Oscillations of a Magnetically \\ Charged Black Hole}

\author{Bidyut Hazarika \orcidlink{0009-0007-8817-1945}$^1$}

\email{rs\_bidyuthazarika@dibru.ac.in}

\author{Mrinnoy M.  Gohain \orcidlink{0000-0002-1097-2124}$^1$}
\email{mrinmoygohain19@gmail.com} 

\author{Prabwal Phukon \orcidlink{0000-0002-4465-7974}$^1$,$^2$}
\email{prabwal@dibru.ac.in}

	\affiliation{$^1$Department of Physics, Dibrugarh University, Dibrugarh, Assam, 786004.\\$^2$Theoretical Physics Division, Centre for Atmospheric Studies, Dibrugarh University, Dibrugarh, Assam, 786004.\\}

%\date{}
\begin{abstract}
In this work, we explore the influence of nonlinear electrodynamics (NED) on the quasi-periodic oscillations (QPOs) of a magnetic charged black hole by analyzing the motion of test particles and their epicyclic frequencies. Starting from the effective potential, angular momentum, and energy of circular orbits, we examine how the NED parameter \( b \) alters the orbital dynamics. We find that as \( b \) increases, the system transitions smoothly from the Reissner–Nordström (RN) regime towards  the Schwarzschild profile, with observable changes in the innermost stable circular orbit (ISCO) and Keplerian frequencies. We further investigate the variation in the radii of QPOs with respect to the NED parameter \( b \) by employing the PR, RP, WD, and ER models. We also perform  Markov Chain Monte Carlo (MCMC) analysis using observational QPO data from a diverse set of black hole sources spanning stellar-mass, intermediate-mass, and supermassive regimes. The MCMC results yield consistent constraints on the parameter \( b \) across all mass regimes, indicating that NED effects leave a distinguishable signature on the QPO structure of a charged black hole. 
\end{abstract}

%\pacs{}

\maketitle                                                                      

\section{Introduction}
Black holes represent some of the most mysterious and fascinating outcomes predicted by General Relativity (GR), underscoring the theory’s profound implications for our understanding of gravity. Since Einstein introduced GR in 1915, it has provided a robust theoretical foundation for describing spacetime curvature. One of the most compelling confirmations of GR came with the observation of gravitational waves from merging black holes by the LIGO collaboration, marking a monumental milestone in gravitational physics \cite{ligo}. This was soon complemented by the Event Horizon Telescope’s (EHT) unprecedented imaging of supermassive black holes  first in the galaxy M87, and later Sagittarius A* (SgrA*) at the heart of the Milky Way \cite{m87a, m87b, m87c, m87d, m87e, m87f}. These groundbreaking images revealed a central shadow encircled by a luminous photon ring, the structure of which encodes crucial information about the nature of the black hole and the underlying gravitational framework \cite{Shadow1,Shadow2,Shadow3,Shadow4}. Extensive research indicates that the morphology of black hole shadows and the characteristics of photon spheres offer promising avenues to probe and constrain deviations from GR, making them essential tools for testing alternative theories of gravity \cite{Shadow5,Shadow6,Shadow7,Shadow8,Shadow9,Shadow10,Shadow11,Shadow12}. \\

Magnetic fields are a pervasive feature in astrophysical environments and play a significant role in shaping the dynamics of charged matter around compact objects. In particular, their influence becomes crucial in the vicinity of black holes,    where strong gravitational and electromagnetic fields can significantly affect the motion of test particles. Notably, the interaction between a black hole's external magnetic field and a particle’s dipole moment provides key insights into the dynamics near magnetized compact objects.

Early works, including Wald's \cite{1} analytical solution for electromagnetic fields around a Kerr black hole immersed in a uniform magnetic field, laid the foundation for extensive investigations into magnetized black hole spacetimes. Subsequent studies extended this framework to a variety of magnetic field configurations—such as dipolar and split-monopole fields—and across different spacetime geometries, examining the behavior of neutral and charged particles under such influences. These investigations are particularly important for understanding accretion dynamics, particle acceleration mechanisms, and jet formation.

The study of particle dynamics near black holes plays a crucial role in understanding their physical and geometrical characteristics. Over the years, extensive research has investigated the motion of both massive and massless particles in various parameterized black hole spacetimes\cite{2,3,4,5,6,7,8,9,10,11,12}.  
Orbital and epicyclic frequencies in axially symmetric and stationary spacetimes have been extensively studied, particularly for their role in understanding the dynamics of particles in black hole environments \cite{11}.  Early developments provided exact analytical solutions for geodesics, laying the groundwork for more advanced analyses \cite{13}. Subsequent investigations extended these results to include the motion of charged test particles in spacetimes influenced by both electric and magnetic fields \cite{14,15}.  Notably, recent findings indicate that the combined effects of electric charge and external magnetic fields in Reissner–Nordström spacetime can mimic the behavior of black holes with intrinsic magnetic charge \cite{16}, adding further complexity to the particle dynamics in such settings. \\

Quasiperiodic oscillations (QPOs) observed in the X-ray emissions from black holes and neutron stars have become a key probe in understanding the physics of strong gravitational fields. These oscillations, marked by variations in brightness at nearly regular intervals, are believed to arise from fundamental processes such as accretion disk dynamics and relativistic gravitational effects.   In particular, twin-peak QPOs detected in certain systems have led to extensive investigations aimed at uncovering their origin, often linked to resonant or oscillatory modes within the accretion disk. The need for more refined theoretical models, complemented by higher-precision observations, has become increasingly evident. Since the initial detection of QPOs through spectral and timing analysis in X-ray binaries \cite{17}, the phenomenon has been widely explored in both observational and theoretical contexts. Among various models, those based on the motion of particles in curved spacetime have gained traction, where the oscillatory behavior is attributed to modulations in the trajectories of charged test particles, leading to the structure and evolution of the accretion flow \cite{18,19,20,21,22,23,24,25,26,27,28,29,30,31,32}.
Recent numerical studies have investigated the mechanisms responsible for QPO generation in black hole environments by solving general relativistic hydrodynamic equations \cite{33} in spacetimes like Kerr and hairy black holes. These simulations reveal that plasma perturbations during accretion can lead to the formation of spiral shock waves, which are closely linked to QPO activity \cite{34,35,36}. Similarly, models based on Bondi–Hoyle–Lyttleton accretion show that shock cones formed in strong gravitational fields can produce characteristic QPO frequencies\cite{37,38,39,40,41}. These frameworks have been effective in explaining observed QPOs in sources such as GRS 1915+105 \cite{42}, and also offer predictions for QPO features near supermassive black holes like M87 \cite{43}. The motion of test particles and the associated quasi-periodic oscillations (QPOs) around black holes have been extensively studied in the literature; see, for instance, Refs.~\cite{c1,c2,c3,c4,c5,c6,c7} for a selection of relevant works.In Ref.~\cite{c2}, black holes arising from nonlinear electrodynamics were analyzed from the perspective of observed quasi-periodic oscillations, where the authors specifically considered regular rotating black hole solutions.
\\

Nonlinear electrodynamics (NED) provides a promising framework for constructing regular field configurations in curved spacetime.\cite{44} In contrast to linear Maxwell theory, NED models characterized by gauge-invariant Lagrangians depending on the electromagnetic field invariant \( F = F_{\mu\nu}F^{\mu\nu} \) can exhibit stress-energy tensors with specific symmetries that mimic vacuum behavior under radial boosts. Among these, theories like Born–Infeld electrodynamics have attracted significant attention due to their appearance in low-energy limits of string theory. Moreover, several NED models \cite{45,46,47,48,49} share appealing characteristics with Born–Infeld theory, including finite electric fields at the origin and finite total electrostatic energy.  In the next paragraph we provide a brief overview of the NED black hole we are using in this work.\\

The action describing NED  black holes as presented in \cite{49},  
\begin{equation}\label{eq2}
I = \int d^4x \sqrt{-g} \left( \frac{R}{16\pi G} + \mathcal{F}(\mathcal{L}) \right),
\end{equation}  
where \(G\) is Newton’s gravitational constant. The corresponding NED Lagrangian \cite{49} for these black hole configurations is given by  
\begin{equation}\label{eq3}
\mathcal{L}(\mathcal{F}) = -\frac{\mathcal{F}}{4\pi \cosh^2\left(a (2|\mathcal{F}|)^{1/4}\right)},
\end{equation}  
with \(a\) representing the coupling parameter and \(\mathcal{F} = F^{\mu\nu}F_{\mu\nu}/4\) being the electromagnetic field invariant.\\

We restrict our attention to magnetically charged black holes only, since in the presence of electric charge, NED theories that reproduce Maxwell behavior in the weak-field limit typically lead to singular geometries \cite{44}.  One of the key motivations for choosing the specific form of the Lagrangian that leads to the background solution eq. \eqref{eq14} is to avoid the Laplacian instability issues commonly associated with regular black holes in NED theories. Recent studies \cite{ned1,ned2} have shown that nonsingular black hole solutions within NED frameworks are generally plagued by Laplacian instabilities near the center, arising from a negative squared sound speed in the angular propagation direction. Such instabilities tend to enhance gravitational perturbations and destabilize the regular background geometry. In contrast, the metric  eq. \eqref{eq14} considered here, which features a central singularity at $r=0$,   is therefore free from the angular Laplacian instability problem.
 The background geometry is assumed to be spherically symmetric and described by the line element  
\begin{equation}\label{eq7}
ds^2 = -f(r) dt^2 + \frac{1}{f(r)} dr^2 + r^2 (d\theta^2 + \sin^2\theta\, d\phi^2).
\end{equation}  
The corresponding metric function for this solution takes the form  \cite{48},
\begin{equation}\label{eq14}
f(r) = 1 - \frac{2MG}{r} + \frac{Q^2 G}{b~ r} \tanh\left(\frac{b}{r}\right) ,
\end{equation}  
Here, we define \( b = a\sqrt{Q} \) to simplify the metric function. Since \( b \propto a \), we interpret \( b \) as the characteristic parameter of NED, encapsulating the effects of the NED coupling. \\

Several notable features emerge in the extremal limits of the parameter \( b \), which are particularly significant in the context of this study.  As \( r \to \infty \), asymptotically eq.\ref{eq14} takes the following form:
\begin{equation}
f(r) = 1 - \frac{2MG_N}{r} + \frac{Q^2G}{r^2} - \frac{Q^2 b^2G^3}{ r^4} + \mathcal{O}(r^{-6}) 
\label{inf}
\end{equation}

In the limit \( b \to 0 \), the metric function in Eq.~\ref{inf} simplifies to that of the Reissner–Nordström (RN) solution with a magnetic charge \( Q \). This indicates that for sufficiently small values of \( b \), the geometry closely resembles the RN black hole. 
On the other hand, from Eq.~\ref{eq14}, it is clear that taking the limit \( b \to \infty \) leads the metric to reduce to the Schwarzschild solution. Hence, in the regime of large \( b \), the spacetime behaves like a Schwarzschild black hole. \\
The absence of angular Laplacian instability can be used to put theoretical bounds on \(b\) appearing in the metric \eqref{eq14}. We next compute \(c^2_\Omega\) outside the outer horizon by using the equation \cite{ned1}
\begin{equation}
c_{\Omega}^2 = \frac{\mathcal{L}_{,\mathcal{F}} + 2\mathcal{F} \mathcal{L}_{,\mathcal{F}\mathcal{F}}}{\mathcal{L}_{,\mathcal{F}}}
\label{eqL}
\end{equation}
Eq.\ref{eqL} can be further simplified to  \cite{ned1}
\begin{equation}
c_{\Omega}^2=-\frac{r \left( r^2 f^{(3)}(r) + 2r f''(r) - 2 f'(r) \right)}{2 \left( r^2 f''(r) - 2 f(r) + 2 \right)}
\end{equation}
the expression of $c_{\Omega}^2$ thus obtained  is
\begin{equation}
c_{\Omega}^2=\frac{2 b^2 \tanh ^2\left(\frac{b}{r}\right)-b^2 \text{sech}^2\left(\frac{b}{r}\right)-7 b r \tanh \left(\frac{b}{r}\right)+4 r^2}{2 r \left(2 r-b \tanh \left(\frac{b}{r}\right)\right)}
\end{equation}
\begin{figure}[!h]
\centering
\includegraphics[width=0.95\linewidth]{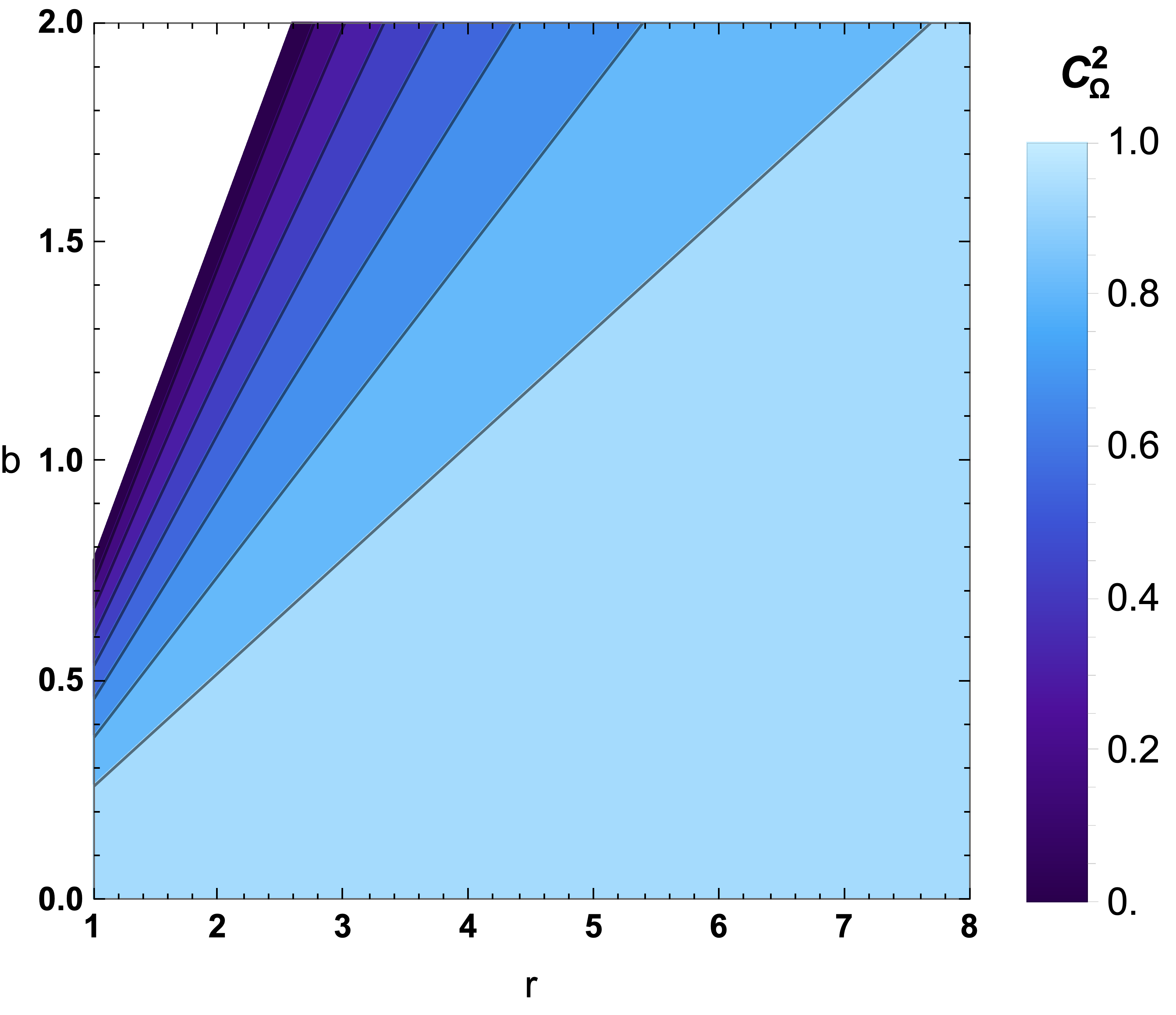}
\vspace{-0.3cm}
\caption{Contour plot for $c_{\Omega}^2$ in the $r-b$ plan. The white portion is the non physical region where $c_{\Omega}^2$ is negative. } 
\label{fig0}
\end{figure}
In order to examine the stability of the black hole solution, we plot the behavior of the angular propagation speed squared, \( c_\Omega^2 \), in the \( r \)-\( b \) plane, as shown in Fig.~\ref{fig0}. The horizontal axis represents the radial coordinate \( r \), while the vertical axis corresponds to the nonlinear electrodynamics parameter \( b \). The color shading indicates that, for a fixed \( b \), increasing \( r \) generally leads to an increase in the value of \( c_\Omega^2 \), with lighter regions corresponding to values approaching unity. As $r \to \infty$, \( c_\Omega^2 \) asymptotically approaches the value \( c_\Omega^2 = 1 \). Consequently,  using Fig.\ref{fig0}, one can identify the physical regions where angular instabilities are absent. The blank region in the figure corresponds to unphysical values, excluded due to the presence of angular instabilities. For instance, in the case of a black hole with an outer horizon radius of 2, the parameter \(b\) must not exceed the threshold value \(b = 1.7721\).More generally,  for black holes with outer horizon radius greater than $2$, values of \(b\) less than  to \(1.7721\) are guaranteed to lie within the physically acceptable region. For values of \(b\) exceeding this threshold, physical viability depends on the specific outer horizon radius and must be assessed accordingly. However,  this bound represents a relatively weak constraint. The method employed here does not offer a rigorous or comprehensive framework for constraining \(b\) based on observational data alone.  It is important to note that the expression for \( c_ \Omega ^2\) in Eq.  \ref{eqL} remains valid both outside and inside the horizon. While \( c_ \Omega ^2\) is positive in the exterior region, there exist a region  inside the horizon where it becomes negative  particularly near \( r = 0 \),  signaling a possible Laplacian instability.  Since the Lagrangian in Eq. \ref{eq3} is part of an effective field theory,  where it is expected to be valid only below a certain energy scale,  typically outside the horizon. A more complete description would require a UV-complete theory that resolves the high-energy behavior near the central singularity.\\

The motivation behind this work, is  to investigate whether the characteristic behavior of the metric observed in the extremal limits of the NED parameter \( b \) also manifests in the context of quasi-periodic oscillations (QPOs). Specifically, we seek to understand whether similar trends emerge in the QPO frequencies as \( b \to 0 \) and \( b \to \infty \), and how the NED parameter influences this behavior. Our central objective is to identify the potential "signatures" of nonlinear electrodynamics encoded in the QPO profiles by addressing these questions.\\

In this study, we explore the motion of neutral test particles in the vicinity of a static, spherically symmetric magnetically charged black hole solution that arises from nonlinear electrodynamics (NED) coupled to general relativity. Our primary goal is to identify the imprints of the NED parameter \( b \) on the quasi-periodic oscillations (QPOs) associated with magnetically charged black holes. To this end, we analyze the orbital properties of test particles,  focusing on the effective potential, angular momentum, and energy of stable circular orbits and investigate how these quantities evolve as a function of \( b \). Our findings indicate a continuous transition in the spacetime geometry from the Reissner–Nordström (RN) regime to a Schwarzschild-like profile as \( b \) increases, accompanied by noticeable changes in the location of the innermost stable circular orbit (ISCO) and the corresponding Keplerian frequencies.
To further probe the influence of the NED parameter, we examine the behavior of QPO radii under different phenomenological models, including the Relativistic Precession (RP), Warped Disk (WD), and Epicyclic Resonance (ER) models. Additionally, we carry out a Markov Chain Monte Carlo (MCMC) analysis using observational QPO data from various black hole sources, encompassing stellar-mass, intermediate-mass, and supermassive regimes. The resulting posterior distributions consistently constrain the parameter \( b \), suggesting that the effects of nonlinear electrodynamics leave a measurable imprint on the QPO spectrum of charged black holes across different mass scales.

\section{Particle Dynamics around NED black holes} 
\label{null}
\subsection{Equations of Motion}
In this section, we explore the motion of electrically neutral test particles in the vicinity of a charged black hole described by nonlinear electrodynamics (NED). The dynamics of these test particles are governed by the following Lagrangian:
\begin{align}
    L_p = \frac{1}{2} m g_{\mu\nu} \dot{x}^\mu \dot{x}^\nu,
\end{align}
where $m$ denotes the mass of the particle, and the dot represents differentiation with respect to the proper time $\tau$.  It is crucial to note that \( x^\mu(\tau) \) characterizes the worldline of the particle, parametrized by the proper time \( \tau \), while the particle's four-velocity is defined as \( u^\mu = \frac{dx^\mu}{d\tau} \).

In a spherically symmetric spacetime, there exist two Killing vectors associated with time-translation and rotational invariance of spacetime, given by \( \xi^\mu = (1, 0, 0, 0) \) and \( \eta^\mu = (0, 0, 0, 1) \), respectively. Hence the constant of motion are  correspond to the total energy \( E \) and angular momentum \( L \) of the test particle, which can be formulated as:

\begin{align}
\mathcal{E}&=-g_{tt}  ~\Dot{t}, \nonumber\\
\mathcal{L}&=g_{\phi \phi}~ \Dot{\phi}.
\label{energy}
\end{align}
In Eq.~(\ref{energy}), the symbols $\mathcal{E}$ and $\mathcal{L}$ represent the energy and angular momentum per unit mass, respectively. The equation of motion for the test particle can be derived using the normalization condition:

\begin{align}\label{normal}
    g_{\mu\nu} u^\mu u^\nu = \delta,
\end{align}

where $\delta = 0$ and $\delta = \pm 1$ correspond to geodesic motion for massless and massive particles, respectively. Specifically, $\delta = +1$ is associated with spacelike geodesics, while $\delta = -1$ corresponds to timelike geodesics. For massive particles, the motion is governed by timelike geodesics of spacetime, and the corresponding equations can be obtained by employing Eq.~(\ref{normal}).

By considering Eqs.~(\ref{energy}) and (\ref{normal}), the equation of motion at a constant
plane can be expressed in the following form:

\begin{align}
    \dot{r}^2=\mathcal{E}+g_{tt}\Big(1+\frac{{\mathcal{L}^2}}{r^2 }\Big),\nonumber\\
      \dot{r}^2=\mathcal{E}+g_{tt}\Big(1+\frac{{\mathcal{L}^2}}{r^2 }\Big).
\end{align}
In a static and spherically symmetric spacetime, if a particle begins its motion in the equatorial plane, it will continue to move within this plane throughout its trajectory. By restricting the motion to the equatorial plane, where $\theta = \frac{\pi}{2}$ and $\dot{\theta} = 0$, the radial equation of motion can be written as:

\begin{align}
    \dot{r}^2 = \mathcal{E}^2 - V_\text{eff},
\end{align}
Now, applying standard conditions for circular motion, 
$\dot{r}=0$, and $\ddot r=0$,  we get the following equations
\begin{align}
    \dot{r}=0,\nonumber\\
     V_\text{eff}=\mathcal{E}^2
     \label{cosnt}
\end{align}\\

\begin{figure}[!h]
\centering
\includegraphics[width=0.95\linewidth]{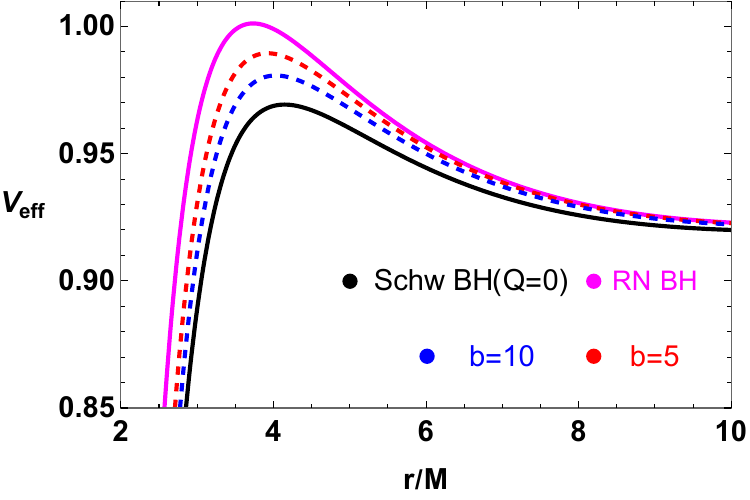}
\vspace{-0.3cm}
\caption{Radial dependence of effective potential for the different values of b and Q.Here we have considered Q=0.5 } 
\label{fig1}
\end{figure}
\vspace{0.5cm}
where the effective potential governing the radial motion in equatorial plane  is given by:

\begin{align}
    V_\text{eff} = f(r) \left(1 + \frac{\mathcal{L}^2}{r^2}\right).
\end{align}
\begin{figure*}[!t]
\centering
\includegraphics[width=0.32\linewidth]{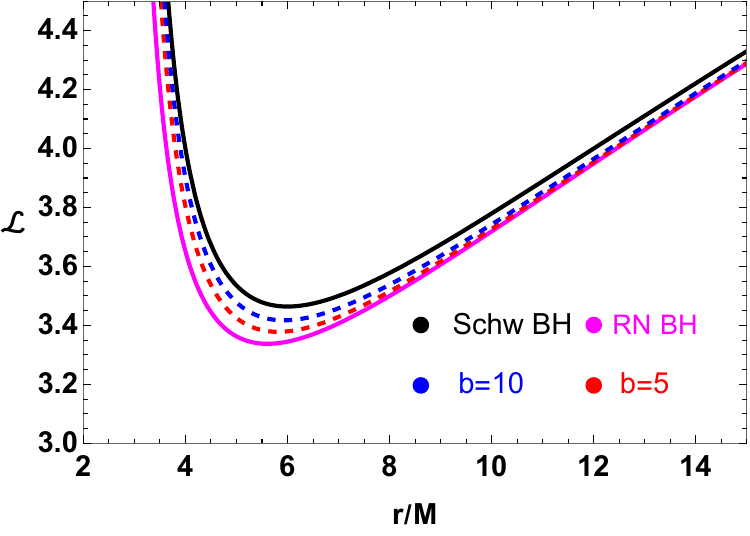}  % Adjust width to fit three figures
\includegraphics[width=0.32\linewidth]{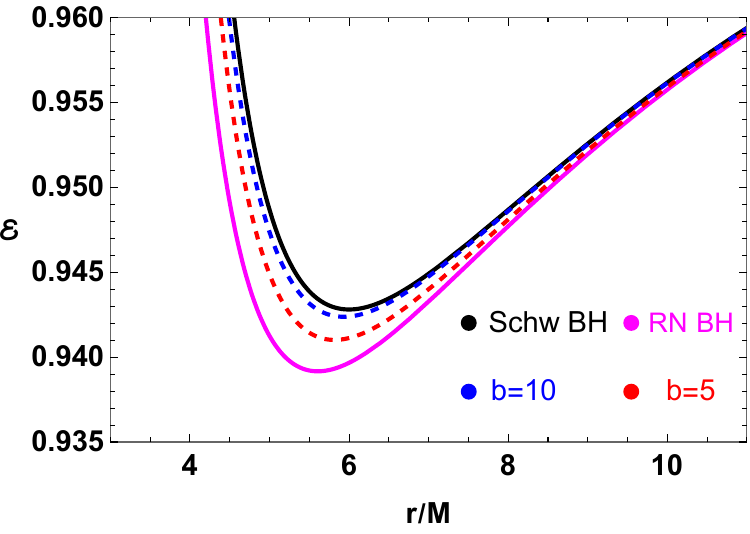}
\includegraphics[width=0.32\linewidth]{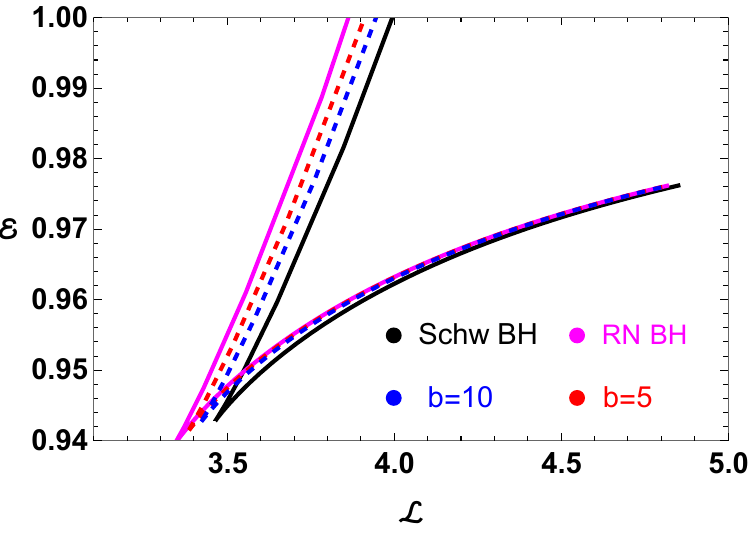}
\vspace{-0.3cm}
\caption{The radial dependence of specific angular momentum and energy  for circular orbits for different values of NED parameter b.  Here, we have considered $Q = 0.5$.}
\label{fig2}
\end{figure*}
Figure \ref{fig1} depicts the radial dependence of the effective potential for a charge less test particles, showing how it varies with the parameter $b$ of the nonlinear electrodynamics (NED) black hole. The figure also compares these results with those for the Reissner-Nordström (RN) black hole and the Schwarzschild black hole(Q=0). For a neutral particle, when the charge parameter of the black hole $Q$ is fixed, increasing the value of $b$ leads to a decrease in the maximum value of the effective potential compared to the RN black hole. As $b$ increases further, the effective potential approaches the profile observed in the Schwarzschild black hole case where charge is zero. The minima in the effective potential correspond to stable circular orbits, where a particle can remain in equilibrium without drifting away. In contrast, the maxima represent unstable circular orbits, where small perturbations can cause the particle to move away from the orbit. The position and depth of these extrema are influenced by the black hole’s parameters, specifically the charge $Q$ and the NED parameter $b$, which govern the spacetime geometry and the particle dynamics.\\

Next, using the expressions Eq.(\ref{cosnt}) , we derive expressions for the specific angular momentum and the specific energy for circular orbits  in the following form:
\begin{align}
    \mathcal{L}=\frac{r^2 \left(2 b M r-Q^2 r \tanh \left(\frac{b}{r}\right)-b Q^2 \text{sech}^2\left(\frac{b}{r}\right)\right)}{2 b r (r-3 M)+3 Q^2 r \tanh \left(\frac{b}{r}\right)+b Q^2 \text{sech}^2\left(\frac{b}{r}\right)},
\end{align}
\begin{align}
    \mathcal{E}=\frac{2 \left(b (r-2 M)+Q^2 \tanh \left(\frac{b}{r}\right)\right)^2}{b \left(2 b r (r-3 M)+3 Q^2 r \tanh \left(\frac{b}{r}\right)+b Q^2 \text{sech}^2\left(\frac{b}{r}\right)\right)}.
\end{align}
Figure \ref{fig2} presents the radial dependence of specific angular momentum $\mathcal{L}$ and energy $\mathcal{E}$ for circular orbits in different black hole spacetimes, considering the influence of the nonlinear electrodynamics (NED) parameter $b$. 
The left panel shows the variation of the specific angular momentum $\mathcal{L}$ with the radial coordinate $r/M$. The solid black line represents the Schwarzschild black hole (Schw BH), while the magenta solid  line corresponds to the Reissner-Nordström black hole (RN BH). The blue and red dashed lines depict modifications due to the NED parameter for $b = 10$ and $b = 5$, respectively. As we increase the value of $b$ we observe a decrease in the minimum value of the  angular momentum  compared to the RN black hole. As $b$ increases further, the  angular momentum approaches the profile observed in the Schwarzschild black hole case . The middle panel illustrates the energy $\mathcal{E}$ of circular orbits as a function of $r/M$. Similar color coding is used as in the left panel. The energy profile exhibits a minimum, indicating the most bound orbit. Compared to the Schwarzschild and RN BH cases, the NED parameter modifies the energy required for stable orbits, affecting the depth and location of the minimum in the same pattern as we observe in case of $\mathcal{L}$ vs $r/M$ plot.  The right panel shows the relationship between $\mathcal{L}$ and $\mathcal{E}$, providing insights into the stability of orbits and their dependency on the NED parameter. It can be concluded, for higher values of $b$,  both the specific angular momentum $\mathcal{L}$ and energy $\mathcal{E}$ of circular orbits decrease, indicating that the orbits become more bound.   Regarding stability, since lower angular momentum and energy imply that the particle requires less effort to remain in orbit, the stable circular orbits tend to shift outward. However, if the reduction in $\mathcal{L}$ and $\mathcal{E}$ is significant, it may also lead to a decrease in the size of the stable region, potentially making certain orbits more prone to instabilities.The differences between the Schwarzschild, RN, and NED-modified cases are evident, with the latter showing deviations due to the parameter $b$.

\subsection{Innermost stable circular orbits (ISCO)}\label{secIII}

Solving the condition $V_{\text{eff}} = 0$ with respect to $r$ allows one to determine the locations where the effective potential exhibits extremal behavior. Stable circular orbits correspond to minima of the effective potential, i.e., when $\partial_r^2 V_{\text{eff}}(r) > 0$, whereas orbits are unstable if $\partial_r^2 V_{\text{eff}}(r) < 0$. The innermost stable circular orbit (ISCO) is identified by the condition $\partial_r^2 V_{\text{eff}}(r_{\text{ISCO}}) = 0$ However solving the equations in case of NED black hole is not straight forward due to involvement of complex hyperbolic terms. We solve the equations numerically and plots the resultant ISCO radius with respect to the parameter $b.$\\

\begin{figure}[!h]
\centering
\includegraphics[width=0.95\linewidth]{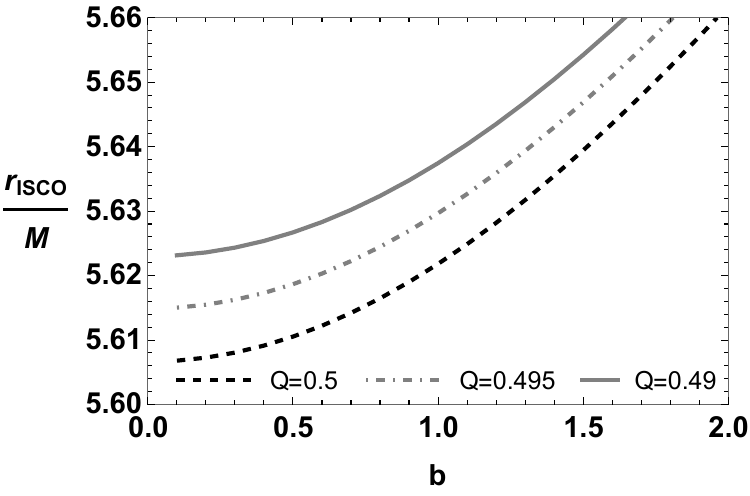}
\vspace{-0.3cm}
\caption{Radius of the ISCO as a function of  parameter $b$ in NED black hole } 
\label{fig3}
\end{figure}
In Fig. \ref{fig3},  we depict the behavior of the ISCO radius as a function of $b$ for different values of the electric charge $Q$. It is evident that for all values of $Q$, the ISCO radius increases monotonically with increasing $b$. This behavior implies that the presence of NED effects pushes the ISCO outward from the black hole. Furthermore, for fixed values of $b$, the ISCO radius tends to exhibit an increasingly linear behavior as the charge parameter $Q$ increases. This trend is expected, as in the limit $Q \to 0$, the ISCO radius per unit mass asymptotically approaches a constant value, yielding a horizontal line at $\frac{r_{\text{ISCO}}}{M} = 6$, which corresponds to the Schwarzschild case.
\section{Fundamental frequencies}
In this section, we compute the fundamental frequencies that characterize the motion of a particle in the vicinity of a NED black hole.  In particular, we focus on the Keplerian frequency, along with the radial and vertical epicyclic frequencies associated with perturbed circular orbits. 
\subsection{Keplerian frequencies}
The angular velocity of a test particle revolving around a black hole, as perceived by a distant observer, is termed the orbital or Keplerian frequency, denoted by $\Omega_\phi$. It is given by the relation $\Omega_\phi = \frac{d\phi}{dt}$. Utilizing this definition, one can derive the general expression for the orbital frequency in a static, spherically symmetric spacetime~\cite{50} as
\begin{align}
    \Omega_\phi = \sqrt{\frac{-\partial_r g_{tt}}{\partial_r g_{\phi\phi}}} = \sqrt{\frac{f'(r)}{2r}}.
\end{align}
For black holes influenced by nonlinear electrodynamics (NED), this expression modifies to:
\begin{align}
\Omega_\phi = \sqrt{\frac{M}{r^3} - \frac{Q^2 \left(r \tanh \left(\frac{b}{r}\right) + b\, \text{sech}^2\left(\frac{b}{r}\right)\right)}{2 b r^4}}.
\end{align}
If we substitute $Q=0$ or $b\to \infty$ limit, we get the same angular velocity as the pure Schwarzschild case \cite{c7} which is :
\begin{align}
\Omega_\phi = \sqrt{\frac{M}{r^3}}.
\end{align}
Again in the $b\to 0$,  limit the angular velocity reduced to a expression same as the RN case :
\begin{align}
\Omega_\phi = \sqrt{\frac{M}{r^3}-\frac{Q^2}{r^4}}.
\end{align}
To convert the angular frequency into physical frequency in units of Hertz (Hz), we employ the following relation:
\begin{align}
    \nu_\phi = \frac{c^3}{2\pi G M} \cdot \sqrt{\frac{M}{r^3} - \frac{Q^2 \left(r \tanh \left(\frac{b}{r}\right) + b\, \text{sech}^2\left(\frac{b}{r}\right)\right)}{2 b r^4}}.
\end{align}
\begin{figure}[!h]
\centering
\includegraphics[width=0.95\linewidth]{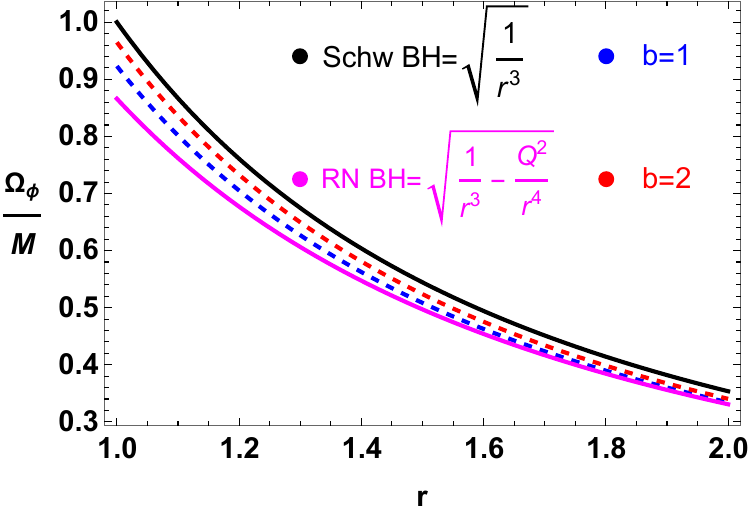}
\vspace{-0.3cm}
\caption{Comparison of angular frequency of NED black hole with Schwarzchild and RN black hole case} 
\label{fig4}
\end{figure}
Fig. \ref{fig4} illustrates the variation of the Keplerian frequency $\Omega_\phi/M$ as a function of radial coordinate $r$. The black solid curve corresponds to the Schwarzschild black hole,  while the magenta curve represents the Reissner–Nordström (RN) black hole.  The blue and red dashed curves depict the behavior for NED black holes with different values of the parameter $b = 1$ and $b = 2$, respectively. It is evident that the orbital frequency decreases with increasing $r$, and the inclusion of charge and nonlinear electrodynamics effects leads to a reduction in $\Omega_\phi$ compared to the Schwarzschild case. Additionally, larger values of $b$ tend to lower the orbital frequency further, highlighting the influence of the NED parameter on the dynamics of test particles.  Consistent with previous studies, we observe that as the parameter $b$ increases, the angular frequency profile gradually converges toward that of the Schwarzschild black hole and as the $b$ decreases, the angular frequency profile gradually converges toward that of the RN  black hole.

\begin{figure*}[t!]
\centering
\includegraphics[width=0.3\linewidth]{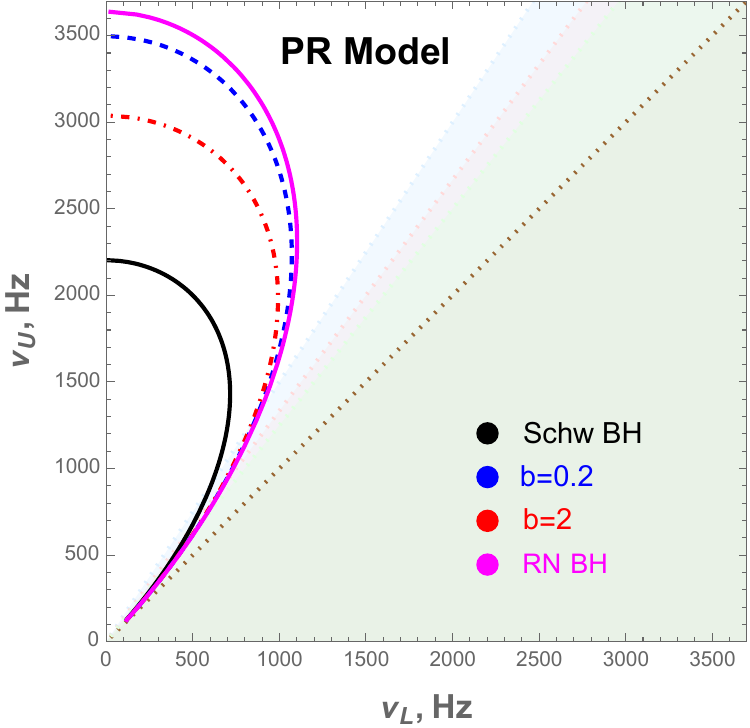}
\includegraphics[width=0.3\linewidth]{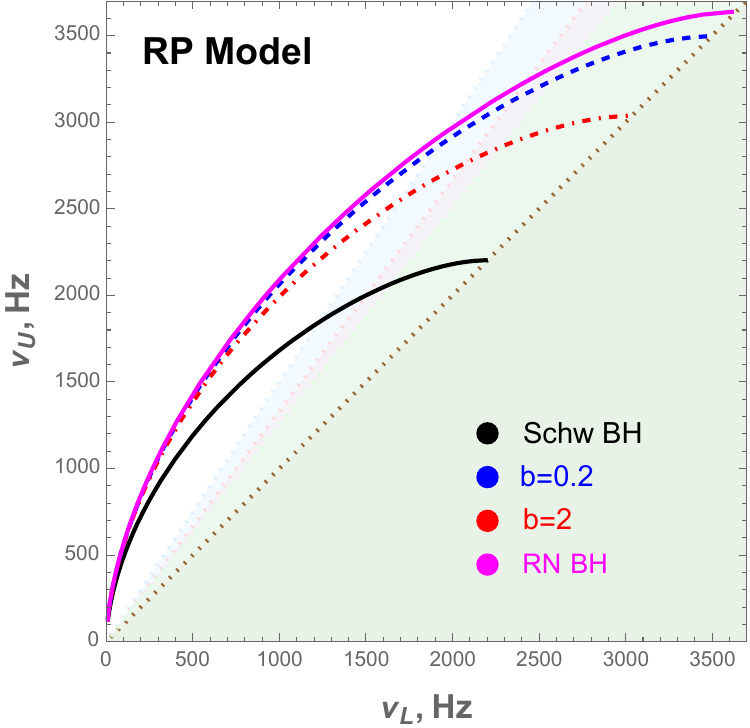}
\includegraphics[width=0.3\linewidth]{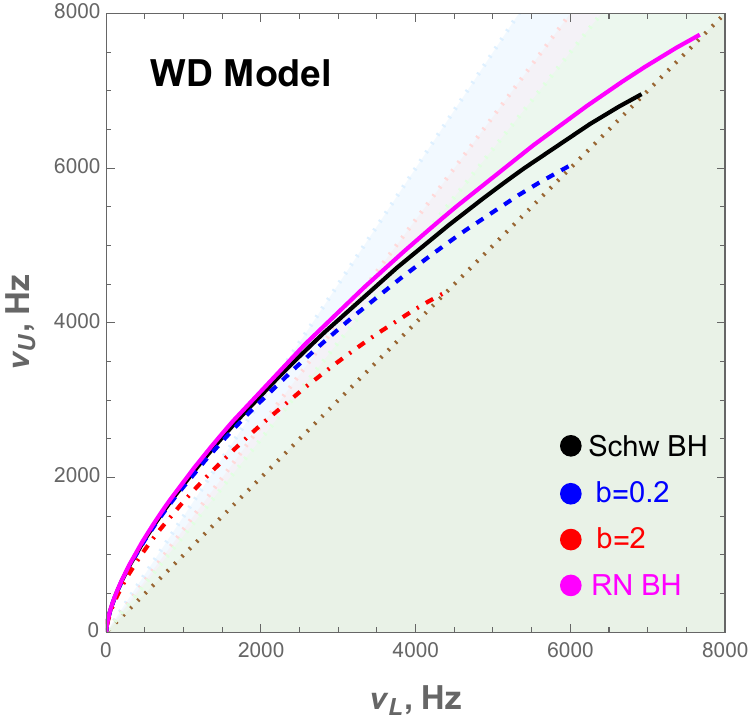}
\includegraphics[width=0.3\linewidth]{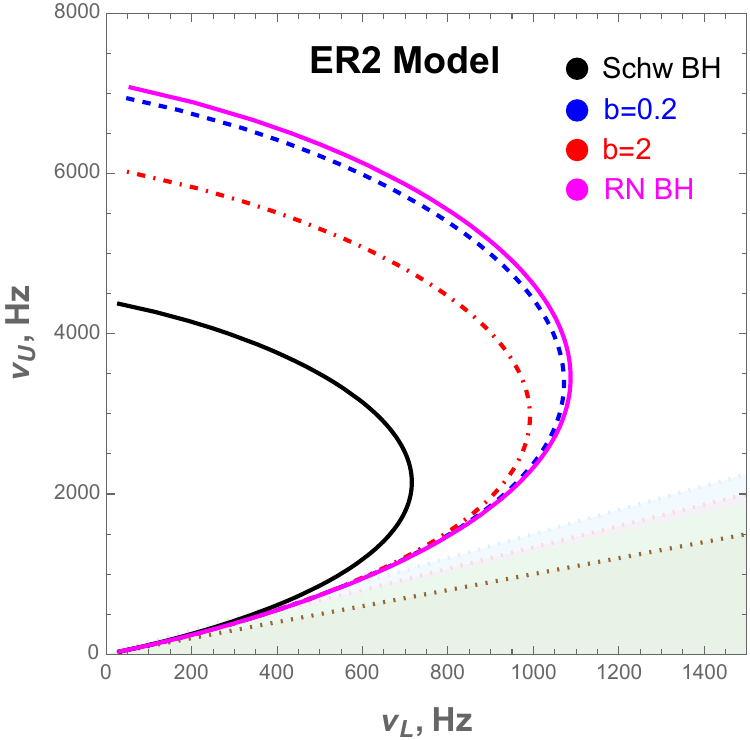}
\includegraphics[width=0.3\linewidth]{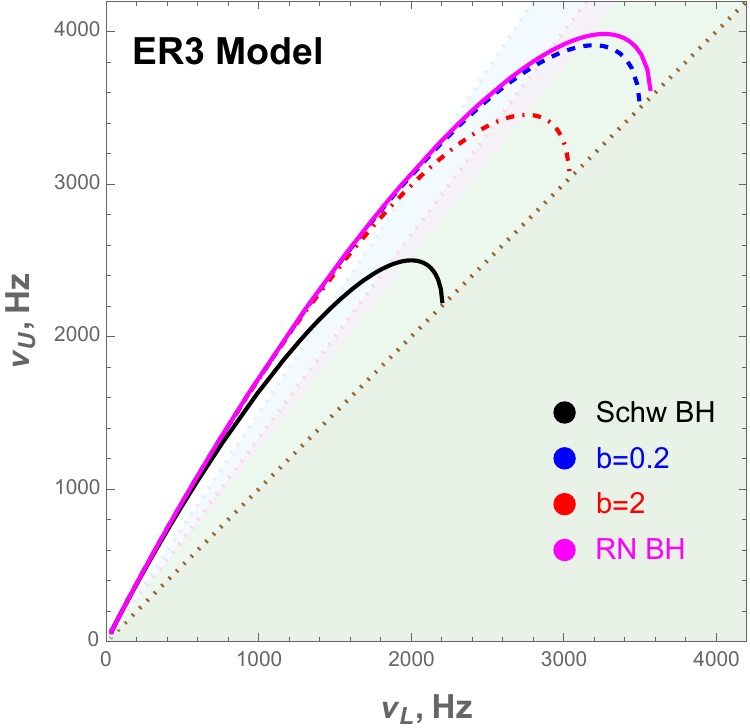}
\includegraphics[width=0.3\linewidth]{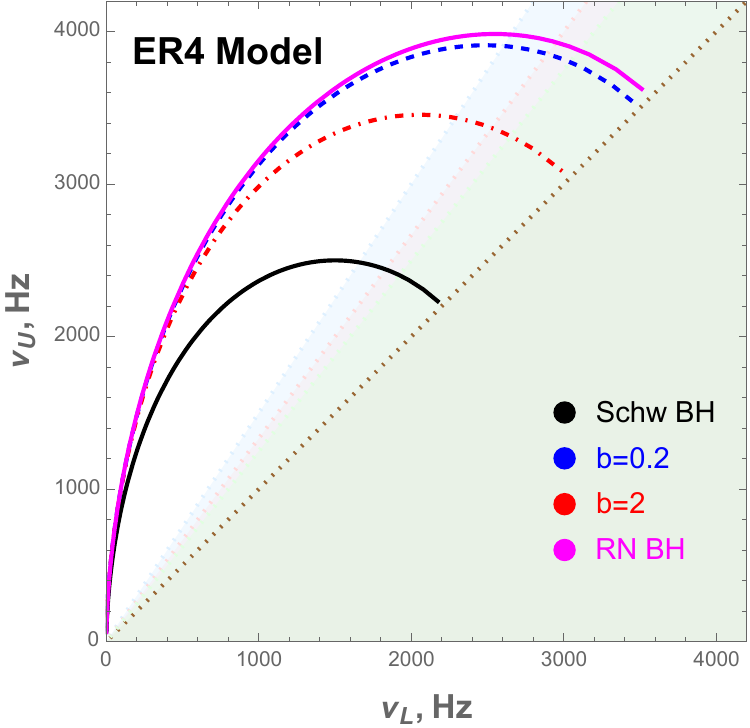}
\caption{Relations between the upper frequency $\nu_U$ and lower frequency $\nu_L$ of twin-peak QPOs for the PR, RP, WD, and ER2–ER4 models in the background of Schwarzschild, RN, and NED black holes. The curves are plotted for different values of the deviation parameter $b$, with $Q = 0.5$ for the RN black hole and $Q = 0$ for the Schwarzschild case.} 
\label{fig5}
\end{figure*}
\subsection{Harmonic oscillations}

In this subsection, we examine the fundamental frequencies associated with the oscillatory motion of test particles orbiting a NED  black hole. These characteristic frequencies, specifically the radial and vertical (or latitudinal) components, can be obtained by introducing small perturbations about the equilibrium circular orbit, i.e., $r \rightarrow r_0 + \delta r$ and $\theta \rightarrow \theta_0 + \delta \theta$. The effective potential $V_{\text{eff}}(r, \theta)$ can be expanded in a Taylor series around the circular orbit $(r_0, \theta_0)$ as follows:
\begin{align}
V_{\text{eff}}(r, \theta) &= V_{\text{eff}}(r_0, \theta_0) + \delta r \left. \frac{\partial V_{\text{eff}}}{\partial r} \right|_{r_0, \theta_0} + \delta \theta \left. \frac{\partial V_{\text{eff}}}{\partial \theta} \right|_{r_0, \theta_0} \nonumber \\
&+ \frac{1}{2} \delta r^2 \left. \frac{\partial^2 V_{\text{eff}}}{\partial r^2} \right|_{r_0, \theta_0} + \frac{1}{2} \delta \theta^2 \left. \frac{\partial^2 V_{\text{eff}}}{\partial \theta^2} \right|_{r_0, \theta_0} \nonumber \\
&+ \delta r \delta \theta \left. \frac{\partial^2 V_{\text{eff}}}{\partial r \partial \theta} \right|_{r_0, \theta_0} + \mathcal{O}(\delta r^3, \delta \theta^3).
\label{exp}
\end{align}
By applying the conditions for circular orbits and stability, only the second-order derivatives of the effective potential contribute, leading to harmonic oscillator equations  in the equatorial plane for the radial and vertical perturbations, observable by a distant observer as \cite{Bambi2017book}:
\begin{align}
    \frac{d^2\delta r}{dt^2}+\Omega^2_r\delta r=0,\  \frac{d^2\delta\theta }{dt^2}+\Omega^2_\theta \delta\theta=0,
\end{align}
where
\begin{align}
 \Omega_r^2=-\frac{1}{2g_{rr}\Dot{t}^2}\partial_r^2 V_\text{eff}(r,\theta)\Big\arrowvert_{\theta=\pi/2},
\end{align}
\begin{align}
        \Omega_\theta^2=-\frac{1}{2g_{\theta\theta}\Dot{t}^2}\partial_\theta^2 V_\text{eff}(r,\theta)\Big\arrowvert_{\theta=\pi/2},
\end{align}
are the frequencies of the radial and vertical oscillations, respectively. 
For NED black holes the expressions for the frequencies of the radial and vertical oscillations are :

\begin{align}
 \Omega_r^2=\text{See appendix \ref{A}},
\end{align}
\begin{align}
        \Omega_\theta^2= \Omega_\phi^2=\frac{M}{r^3}-\frac{Q^2 \left(r \tanh \left(\frac{b}{r}\right)+b \text{sech}^2\left(\frac{b}{r}\right)\right)}{2 b r^4}
\end{align}
To convert these frequencies into physical frequencies in units of Hertz (Hz), we employ the following relation:
\begin{align}
    \nu_i = \frac{c^3}{2\pi G M} \cdot \Omega_i
\end{align}

\section{QPO Models and QPO orbits}
\subsection{QPO Models}
\begin{figure*}[t!]
\centering
\includegraphics[width=0.3\linewidth]{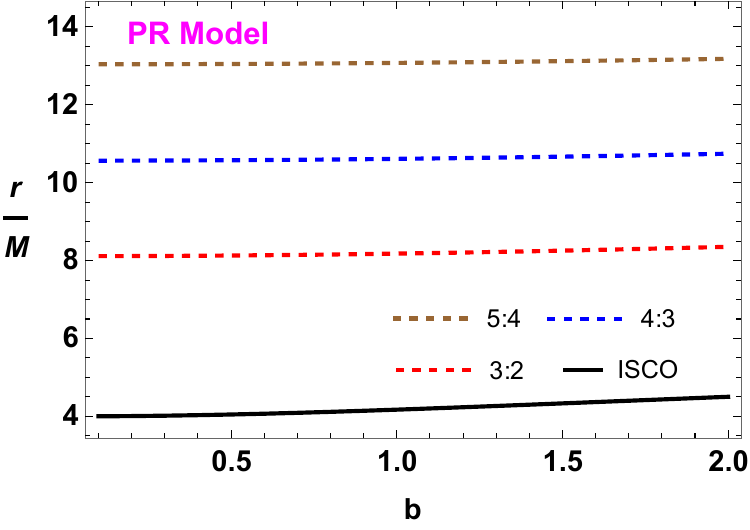}
\includegraphics[width=0.3\linewidth]{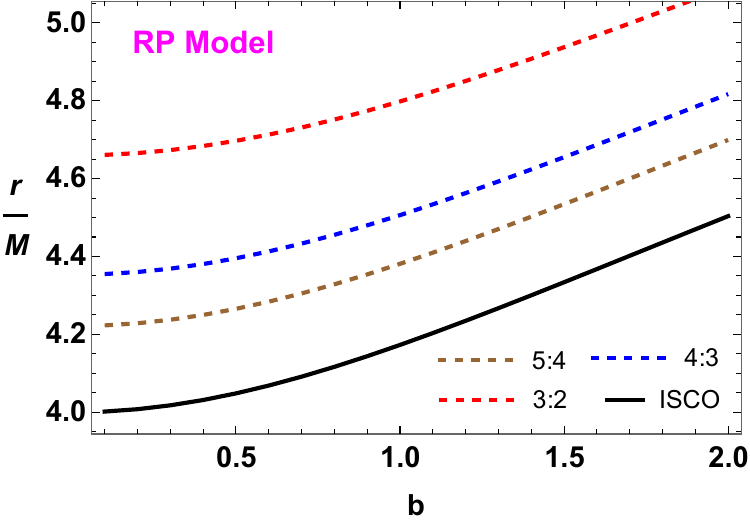}
\includegraphics[width=0.3\linewidth]{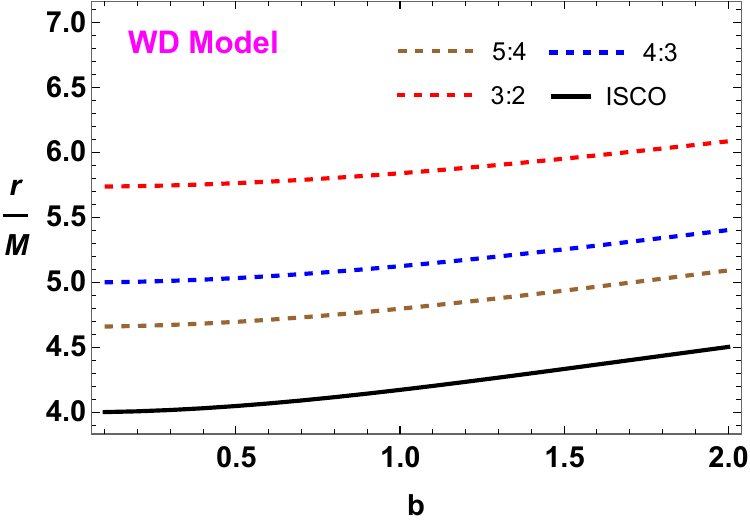}
\includegraphics[width=0.3\linewidth]{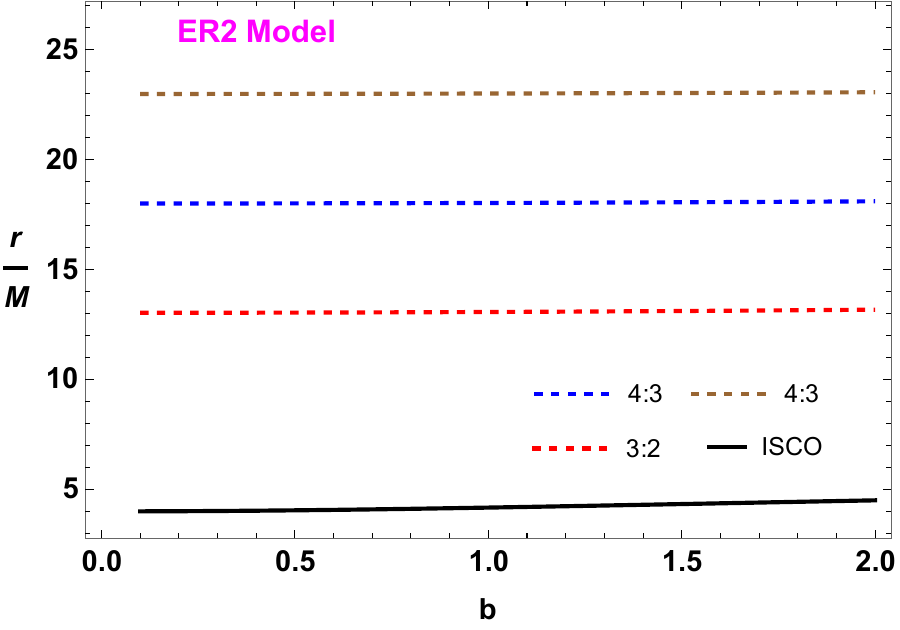}
\includegraphics[width=0.3\linewidth]{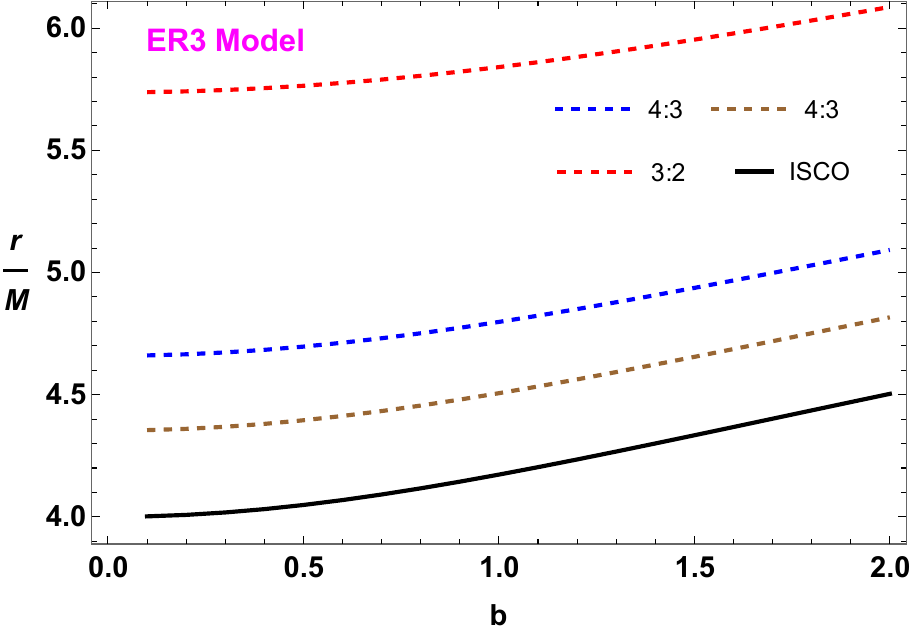}
\includegraphics[width=0.3\linewidth]{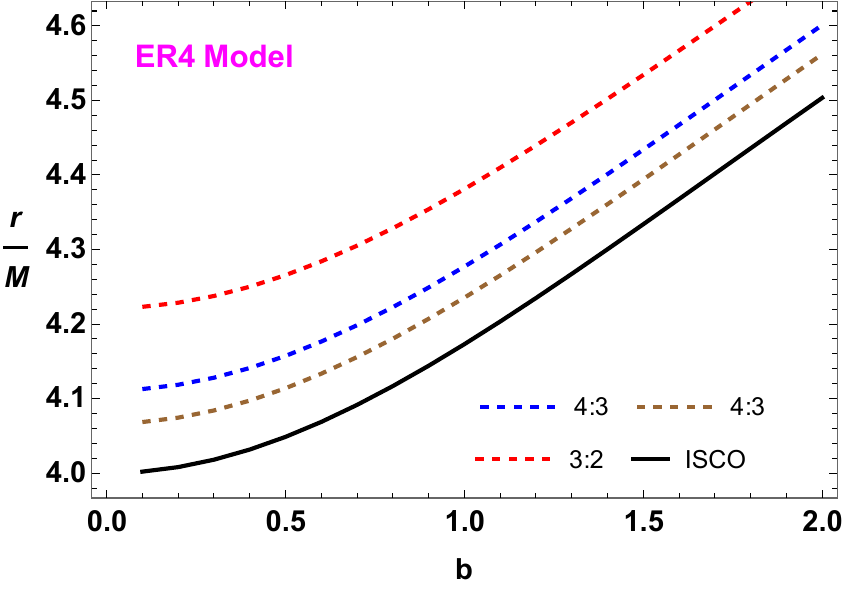}
\caption{ Radius of QPO orbits as a function of the NED  parameter $b$ in PR,  RP, WD, and ER2-4 models} 
\label{fig6}
\end{figure*}
In this section, we investigate the behavior of twin-peak quasi-periodic oscillations (QPOs) in the context of NED black hole, comparing the results with those obtained for the Schwarzschild and RN black hole. The upper ($\nu_U$) and lower ($\nu_L$) QPO frequencies are expressed as functions of the radial coordinate and black hole parameters, in accordance with various established QPO models\cite{51}.  The analysis includes the following QPO models\cite{51}. :
\begin{itemize}
 \item \textbf{Parametric resonance (PR) models:} The observed \(3:2\) ratio in the twin-peak high-frequency QPOs (HFQPOs) in black hole and neutron star systems has motivated the idea that these oscillations may result from resonances between different modes of accretion disk motion~\cite{pr1,pr2,pr3}. In the simplest approach, small radial and vertical perturbations around circular, equatorial particle orbits are treated as independent harmonic oscillations, characterized by the radial and vertical epicyclic frequencies, \(\nu_{r}\) and \(\nu_{\theta}\), respectively. 
The parametric resonance model proposes that the stronger radial oscillations can parametrically excite vertical oscillations, given that perturbations in the radial direction (\(\delta r\)) are typically larger than those in the vertical direction (\(\delta \theta\)) in thin accretion disks. In this framework, vertical oscillations can be excited when the ratio of the radial to vertical epicyclic frequencies satisfies a simple resonance condition, \(\nu_{r}/\nu_{\theta} = 2/n\) for some positive integer \(n\).  In the context of rotating black holes, where typically \(\nu_{\theta} > \nu_{r}\), the strongest resonance occurs for \(n=3\), naturally leading to a \(3:2\) ratio between the oscillation frequencies. In this model the upper and lower frequencies are given by
$\nu_U = \nu_\theta $, $\nu_L = \nu_r$
    \item \textbf{Relativistic Precession (RP) model:} In the relativistic precession (RP) model, quasi-periodic oscillations (QPOs) observed in X-ray binaries are interpreted as a natural outcome of the motion of matter in the curved spacetime around a compact object. In this framework, blobs of plasma in the accretion disk are assumed to follow slightly eccentric and tilted geodesic orbits near the black hole. These small perturbations away from circular motion give rise to characteristic frequencies associated with orbital dynamics.
According to the RP model, the high-frequency QPOs arise from fundamental coordinate frequencies of geodesic motion. The upper kHz QPO is identified with the orbital (Keplerian) frequency, \(\nu_{U} = \nu_{\phi}\), while the lower kHz QPO is associated with the periastron precession frequency, given by \(\nu_{L} = \nu_{\phi} - \nu_{r}\), where \(\nu_{r}\) is the radial epicyclic frequency. Additionally, the low-frequency QPOs observed in black hole systems are linked to the nodal precession frequency, \(\nu_{\phi} - \nu_{\theta}\), which captures vertical oscillations due to the frame-dragging effect. Notably, this nodal precession frequency vanishes in the Schwarzschild limit where \(\nu_{\theta} = \nu_{\phi}\).
These frequency identifications are supported by general relativistic corrections to orbital motion, including frame dragging and spacetime curvature, which naturally produce the observed precessional phenomena. Observations of harmonic structures in both neutron star and black hole systems further support this interpretation, especially in cases where even harmonics of the nodal precession frequency dominate the power spectrum \cite{StellaVietri1998, StellaVietri1999, MorsinkStella1999}. The RP model thus offers a simple, geometrically motivated explanation for QPOs, relying solely on relativistic effects rather than invoking resonance mechanisms or strong magnetic fields.

    \item \textbf{Epicyclic Resonance (ER) models:} Assuming a thick accretion disk, resonance conditions define the frequencies as:
    \begin{itemize}
        \item ER2: $\nu_U = 2\nu_\theta - \nu_r$, $\nu_L = \nu_r$,
        \item ER3: $\nu_U = \nu_\theta + \nu_r$, $\nu_L = \nu_\theta$,
        \item ER4: $\nu_U = \nu_\theta + \nu_r$, $\nu_L = \nu_\theta - \nu_r$.
    \end{itemize}
    \item \textbf{Warped Disk (WD) model:} This model, which assumes a somewhat non-standard geometry for the accretion disk~\cite{wd1,wd2}, explains high-frequency QPOs (HFQPOs) as arising from nonlinear resonances between the relativistically deformed, warped disk and various disk oscillation modes. These resonances involve both horizontal and vertical interactions: horizontal resonances can excite both g-mode and p-mode oscillations, while vertical resonances are capable of exciting only g-mode oscillations~\cite{wd2}. The physical origin of these resonances is linked to the non-monotonic behavior of the radial epicyclic frequency as a function of the radial coordinate \(r\)  Within this framework, the upper high-frequency QPO is identified with ~\cite{wd1,wd2},\\
 $\nu_U = 2\nu_\phi - \nu_r$, $\nu_L = 2(\nu_\phi - \nu_r)$
\end{itemize}

Fig.\ref{fig5} illustrates the computed relations between $\nu_U$ and $\nu_L$ for the NED, RN and Schwarzschild black holes under each QPO model for several values of the parameter $b$. In case of RN black hole, the charge is considered to be $Q=0.5$. The diagram includes reference lines with frequency ratios of $3\!:\!2$, $4\!:\!3$, $5\!:\!4$, and $1\!:\!1$. The latter corresponds to cases where both QPO peaks merge, producing a single peak, sometimes referred to as the "QPO graveyard."
\subsection{QPO orbits}
In this subsection, we examine how the  parameter $b$ influence the orbital radii at which QPOs exhibiting frequency ratios such as $3\!:\!2$, $4\!:\!3$, and $5\!:\!4$ may arise across all considered models. These specific radii can be determined by solving the resonance condition
\begin{equation}
\alpha\,\nu_U(M, r, \ell) = \beta\,\nu_L(M, r, \ell), \label{resonance}
\end{equation}
where $\alpha$ and $\beta$ are integers representing the resonant ratio. For the PR, RP, WD, and ER2–ER4 models, this equation can be solved numerically for the radial coordinate $r$ for different value of the parameter $b$. We plot the numerically solutions in Fig.\ref{fig6}.\\

Fig. \ref{fig6} illustrates the variation of the QPO-generating orbital radius, expressed in units of $r/M$, as a function of the NED parameter $b$ for all six QPO models: PR, RP, WD, ER2, ER3, and ER4. Each panel shows the orbital locations corresponding to the resonant frequency ratios $3\!:\!2$, $4\!:\!3$, and $5\!:\!4$, along with the ISCO  radius depicted by a solid black line. It is evident that the ER2 model produces QPOs at significantly larger radii compared to the other models, indicating the highest magnitude of QPO-generating orbits among the six. The WD model follows, while the ER4 model yields the smallest QPO orbital radii. Moreover, for all models, the orbital radius at which the resonance occurs increases monotonically with the parameter $b$,  indicating that as the deviation from standard electrodynamics becomes stronger, the resonance orbits tend to form farther away from the black hole.

{\LARGE
\begin{table*}[htb]
\centering
\begin{tabular}{|c|c|c|c|}
\hline
\textbf{Source} & \textbf{Mass} (in $M_\odot$) & \textbf{Upper Frequency (Hz)} & \textbf{Lower Frequency (Hz)} \\
\hline
GRO J1655$-$40 & $5.4 \pm 0.3$ \cite{t57} & $441 \pm 2$ \cite{t58} & $298 \pm 4$ \cite{t58} \\
\hline
XTE J1550$-$564 & $9.1 \pm 0.61$ \cite{t59} & $276 \pm 3$ & $184 \pm 5$ \\
\hline
GRS 1915$+$105 & $12.4^{+2.0}_{-1.8}$ \cite{t60} & $168 \pm 3$ & $113 \pm 5$ \\
\hline
H 1743$+$322 & $8.0 - 14.07$ \cite{t61,t62,t63} & $242 \pm 3$ & $166 \pm 5$ \\
\hline
M82 X-1 & $415 \pm 63$ \cite{nature} & $5.07 \pm 0.06$\cite{nature}  & $3.32 \pm 0.06$\cite{nature}  \\
\hline
Sgr A$^*$ & $(3.5 - 4.9) \times 10^6$ \cite{t64,t65} & $(1.445 \pm 0.16) \times 10^{-3}$\cite{t66} & $(0.886 \pm 0.04) \times 10^{-3}$\cite{t66} \\
\hline
\end{tabular}
\caption{Observational QPO data for different Black hole sources with estimated mass  \cite{52}}
\label{tab1}
\end{table*}
}
\begin{table*}[htb]
\centering
\setlength{\tabcolsep}{12pt} % wider column spacing
\renewcommand{\arraystretch}{1.4} % taller rows
\begin{tabular}{|c|c|c|c|c|}
\hline
\textbf{Source} & \boldmath{$M$} & \boldmath{$b$} & \boldmath{$Q/M$} & \boldmath{$r/M$} \\
\hline
  \textbf{GRO J1655-40}     & $5.75 \pm 0.37$                  & $1.06^{+0.63}_{-0.94}$ & $0.24^{+0.13}_{-0.20}$ & $5.45^{+0.23}_{-0.27}$ \\
\hline
\textbf{XTE J1550-564}   & $9.9 \pm 1.1$                    & $1.15^{+0.69}_{-0.62}$ & $0.39 \pm 0.19$        & $5.17^{+0.39}_{-0.45}$ \\

\hline
 \textbf{GRS 1915+105}   & $14.37^{+0.57}_{-0.33}$          & $1.38 \pm 0.87$        & $0.28^{+0.15}_{-0.17}$ & $5.60^{+0.12}_{-0.16}$ \\
\hline
\textbf{H 1743+322}     & $12.6 \pm 1.3$                   & $1.35^{+0.76}_{-0.88}$ & $0.42^{+0.24}_{-0.31}$ & $4.76^{+0.38}_{-0.43}$ \\
\hline
\textbf{M82 X-1}  & $407^{+0.80}_{-1.00}$  & $1.008 \pm 0.30$    & $0.084^{+0.041}_{-0.072}$ & $3.0091^{+0.0052}_{-0.0080}$ \\
\hline
\textbf{Sgr A*}  & $(4.17^{+0.39}_{-0.46}) \times 10^6$ & $1.53 \pm 0.85$    & $0.54^{+0.32}_{-0.27}$ & $4.9^{+1.2}_{-1.7}$ \\
\hline

\end{tabular}
\caption{Posterior estimates of the parameters \(M\), \(b\), \(Q/M\), and \(r/M\) obtained from MCMC analysis.}
\label{tab2}
\end{table*}

\section{Monte Carlo Markov chain (MCMC) analysis}

In this section, we perform a Markov Chain Monte Carlo (MCMC) analysis to constrain the parameters of the NED black hole model using observational data from six well-known black hole sources spanning three different mass regimes: GRO J1655–40, XTE J1550–564, GRS 1915+105, H 1743–322, M82 X-1, and Sgr A*. Among these, GRO J1655–40, XTE J1550–564, GRS 1915+105, and H 1743–322 are stellar-mass black holes, M82 X-1 represents an intermediate-mass black hole, and Sgr A* is a supermassive black hole. The selected black holes and their corresponding observational data are summarized in Table \ref{tab1}.  \\
In this work, we adopted the relativistic precession (RP) model as a representative example to demonstrate the application of MCMC techniques in constraining black hole parameters. The choice was not based on any particular preference for the RP model but rather to illustrate the methodology. However, motivated by the fact that the observed QPO frequencies often satisfy the ratio $\nu_U/ \nu_L$ we have choose 3:2 frequency profile. Notably, the parametric resonance (PR) model provides a natural explanation for this observed ratio, and the corresponding results obtained within this framework are therefore of significant importance.\\
 The Bayesian posterior distribution is given by:
\begin{equation}
P(\boldsymbol{\theta} | D, M) = \frac{P(D | \boldsymbol{\theta}, M)\, \pi(\boldsymbol{\theta} | M)}{P(D | M)},
\label{eq:posterior}
\end{equation}
where $\pi(\boldsymbol{\theta})$ denotes the prior distribution for the parameters $\boldsymbol{\theta} = \{M, b,\frac{Q}{M},  \frac{r}{M}\}$, and $P(D | \boldsymbol{\theta}, M)$ is the likelihood function. We assume Gaussian priors for each parameter, defined as:
\begin{equation}
\pi(\theta_i) \propto \exp\left(-\frac{1}{2} \left(\frac{\theta_i - \theta_{0,i}}{\sigma_i} \right)^2\right), \quad \theta_{\mathrm{low},i} < \theta_i < \theta_{\mathrm{high},i},
\end{equation}
where $\theta_{0,i}$ and $\sigma_i$ represent the mean and standard deviation from literature, and the bounds ensure physical viability. 

The likelihood function includes contributions from the upper and lower QPO frequencies:
\begin{equation}
\log \mathcal{L} = \log \mathcal{L}_U + \log \mathcal{L}_L,
\end{equation}
with
\begin{equation}
\log \mathcal{L}_U = -\frac{1}{2} \sum_i \frac{(\nu^{\mathrm{obs}}_{\phi,i} - \nu^{\mathrm{th}}_{\phi,i})^2}{(\sigma^{\mathrm{obs}}_{\phi,i})^2},
\end{equation}
\begin{equation}
\log \mathcal{L}_L = -\frac{1}{2} \sum_i \frac{(\nu^{\mathrm{obs}}_{L,i} - \nu^{\mathrm{th}}_{L,i})^2}{(\sigma^{\mathrm{obs}}_{L,i})^2},
\end{equation}
where $\nu^{\mathrm{obs}}_{\phi,i}$ and $\nu^{\mathrm{obs}}_{L,i}$ represent the observed orbital and lower QPO frequencies, respectively, while $\nu^{\mathrm{th}}_{\phi,i}$ and $\nu^{\mathrm{th}}_{L,i}$ are the corresponding theoretical predictions derived from the RP model.

We use observational QPO data from the six BH systems mentioned above, which are listed in Table~\ref{tab2}. Based on the prior information, we draw $10^5$ samples for each parameter using Gaussian priors, allowing us to thoroughly explore the multidimensional parameter space. The aim is to extract the most probable values of $\{M,b,\frac{Q}{M}, \frac{r}{M}\}$ that are consistent with observations.

Fig.\ref{fig7} presents the corner plots from our MCMC simulations, with the shaded regions representing the 1$\sigma$ (68\%) and 2$\sigma$ (95\%) confidence intervals for the posterior distributions. The inferred black hole masses span across three different mass regimes, from stellar-mass to supermassive black holes, with the results summarized in Table~\ref{tab2}.
For the stellar-mass black holes, we obtain mass estimates consistent with observational constraints: \(M = 5.75 \pm 0.37\,M_\odot\) for GRO J1655–40, \(9.9 \pm 1.1\,M_\odot\) for XTE J1550–564, \(14.37^{+0.57}_{-0.33}\,M_\odot\) for GRS 1915+105, and \(12.6 \pm 1.3\,M_\odot\) for H 1743+322. The corresponding values of the NED parameter \(b\) lie between approximately 1.06 and 1.38, with moderate uncertainties, while the charge-to-mass ratio \(Q/M\) ranges from 0.24 to 0.42. The radius parameter \(r/M\) for these sources falls between 4.76 and 5.60.
The intermediate-mass black hole, M82 X-1, yields a well-constrained mass of \(M = 407^{+0.80}_{-1.00}\,M_\odot\), with \(b = 1.008 \pm 0.30\), a small charge-to-mass ratio of \(Q/M = 0.084^{+0.041}_{-0.072}\), and  \(r/M = 3.0091^{+0.0052}_{-0.0080}\). These values suggest a weakly charged black hole with a mildly nonlinear electrodynamics contribution.
For the supermassive black hole Sgr A*, we find \(M = (4.17^{+0.39}_{-0.46}) \times 10^6\,M_\odot\), and the NED parameter \(b = 1.53 \pm 0.85\). The inferred charge-to-mass ratio is relatively high, \(Q/M = 0.54^{+0.32}_{-0.27}\), while the parameter \(r/M = 4.9^{+1.2}_{-1.7}\) aligns well with the shadow radius constraints from Keck and VLTI observations, which limit \(r_{\text{sh}}/M\) to \(4.55 \lesssim r_{\text{sh}}/M \lesssim 5.22\) at 1$\sigma$ and \(4.21 \lesssim r_{\text{sh}}/M \lesssim 5.56\) at 2$\sigma$.From the MCMC analysis, we observe that the NED parameter \(b\) remains consistently of order unity across all black hole sources, regardless of their mass regime. This indicates a  persistent deviation from standard linear electrodynamics.  The relatively narrow and consistent bounds on \(b\) may indicate  that the nonlinear corrections could be playing a role in shaping the QPO frequencies.  Therefore, the inferred posterior distributions of the NED parameter and related quantities reflect consistent signatures of nonlinear electrodynamics corrections.
\begin{table*}[htbp]
\centering
\renewcommand{\arraystretch}{1.5}
\setlength{\tabcolsep}{13pt}
\caption{Best-fit parameters from various sources for different models.}
\begin{tabular}{|l|p{3cm}|p{3cm}|p{3cm}|p{3cm}|}
\hline
\textbf{Model: PR} & \(\boldsymbol{b}\) & \(\boldsymbol{M~(M_\odot)}\) & \(\boldsymbol{Q/M}\) & \(\boldsymbol{r/M}\) \\
\hline
XTE & \(1.8861^{+1.0421}_{-1.1940}\) & \(5.8488^{+0.8585}_{-1.0501}\) & \(1.2719^{+0.0428}_{-0.0628}\) & \(5.1535^{+0.9175}_{-0.5476}\) \\

\hline
GRO & \(1.6020^{+0.3804}_{-0.5695}\) & \(3.3917^{+0.4720}_{-0.3168}\) & \(1.2738^{+0.0272}_{-0.0359}\) & \(5.5015^{+0.4465}_{-0.5615}\) \\

\hline
H1743+322 & \(0.9402^{+1.0238}_{-0.9033}\) & \(8.3012^{+3.2837}_{-2.4961}\) & \(1.2981^{+0.0365}_{-0.0442}\) & \(4.2702^{+1.5088}_{-1.0977}\) \\

\hline
GRS & \(0.9779^{+1.9184}_{-0.9043}\) & \(11.8981^{+5.9630}_{-4.4741}\) & \(1.2796^{+0.0597}_{-0.1276}\) & \(4.3231^{+2.1537}_{-1.3058}\) \\

\hline
M82 & \(1.2351^{+0.7357}_{-1.0471}\) & \(456.3790^{+5.7845}_{-5.9466}\) & \(1.5664^{+0.0419}_{-0.0402}\) & \(4.8179^{+0.1022}_{-0.0874}\) \\

\hline
Sgr A* & \(1.4162^{+1.4918}_{-1.2961}\) & \((4.17^{+0.68}_{-0.63})\times10^6\) & \(0.4548^{+0.5128}_{-0.4100}\) & \(5.5650^{+2.2774}_{-2.4416}\) \\

\hline
\textbf{Model: ER2} &  &  &  &  \\
\hline
XTE & \(1.6471^{+1.2724}_{-1.2593}\) & \(5.3787^{+2.3937}_{-1.0300}\) & \(1.4118^{+0.0365}_{-0.0491}\) & \(6.1347^{+1.0824}_{-1.5388}\) \\
\hline
GRO & \(1.8283^{+0.7791}_{-0.6655}\) & \(3.9767^{+0.4494}_{-0.9111}\) & \(1.4246^{+0.0154}_{-0.0141}\) & \(5.3369^{+1.1963}_{-0.4820}\) \\
\hline
GRS & \(1.5887^{+1.3890}_{-1.4698}\) & \(15.0332^{+1.8006}_{-3.3066}\) & \(1.3727^{+0.0436}_{-0.0275}\) & \(3.9817^{+0.8969}_{-0.4411}\) \\
\hline
H1743+322 & \(0.8063^{+1.1364}_{-0.7005}\) & \(9.4218^{+1.7401}_{-2.5189}\) & \(1.3949^{+0.0435}_{-0.0358}\) & \(4.2733^{+1.2489}_{-0.5987}\) \\
\hline
M82 & \(1.2692^{+0.7142}_{-1.1989}\) & \(453.4491^{+2.1229}_{-3.0434}\) & \(1.5681^{+0.0151}_{-0.0162}\) & \(4.8406^{+0.1201}_{-0.0933}\) \\
\hline
Sgr A* & \(1.5458^{+1.3618}_{-1.4849}\) & \((4.20^{+0.66}_{-0.68})\times10^6\) & \(0.5055^{+0.4682}_{-0.4790}\) & \(5.0674^{+2.8246}_{-1.9934}\) \\
\hline
\textbf{Model: ER3} &  &  &  &  \\
\hline
XTE & \(0.7536^{+1.5522}_{-0.6663}\) & \(8.3657^{+2.0527}_{-1.4475}\) & \(1.0064^{+0.1758}_{-0.2393}\) & \(5.6649^{+0.9894}_{-0.9907}\) \\
\hline
GRO & \(1.5922^{+0.3744}_{-0.5168}\) & \(4.2085^{+0.6986}_{-0.5372}\) & \(0.7348^{+0.2051}_{-0.3028}\) & \(6.7409^{+0.8138}_{-0.8314}\) \\
\hline
GRS & \(1.1694^{+1.7062}_{-1.1233}\) & \(11.9699^{+4.3126}_{-2.7693}\) & \(0.9174^{+0.2932}_{-0.7791}\) & \(6.3236^{+1.5837}_{-1.4144}\) \\
\hline
H1743+322 & \(1.0187^{+0.9361}_{-0.9321}\) & \(8.9113^{+3.8331}_{-1.8350}\) & \(0.9220^{+0.2254}_{-0.2101}\) & \(5.9343^{+1.3215}_{-1.5612}\) \\
\hline
M82 & \(0.9527^{+1.1038}_{-0.9289}\) & \(401.7948^{+1.3774}_{-1.7098}\) & \(0.0224^{+0.0640}_{-0.0200}\) & \(6.0065^{+0.0046}_{-0.0118}\) \\
\hline
Sgr A* & \(1.0947^{+0.8369}_{-1.0188}\) & \((3.71^{+0.10}_{-0.11})\times10^8\) & \(1.0888^{+0.8561}_{-0.9992}\) & \(5.7085^{+2.1295}_{-2.5866}\) \\
\hline
\textbf{Model: ER4} &  &  &  &  \\
\hline
XTE & \(0.7490^{+2.0054}_{-0.7308}\) & \(8.7348^{+0.6140}_{-0.6145}\) & \(0.8936^{+0.0590}_{-0.0818}\) & \(4.7741^{+0.3101}_{-0.2612}\) \\
\hline
GRO & \(1.0984^{+0.8128}_{-1.0017}\) & \(4.0058^{+0.1410}_{-0.0498}\) & \(0.1424^{+0.2135}_{-0.1302}\) & \(6.1990^{+0.0584}_{-0.1711}\) \\
\hline
GRS & \(0.8998^{+1.8998}_{-0.8639}\) & \(16.9701^{+2.0622}_{-3.9988}\) & \(1.0167^{+0.0551}_{-0.2385}\) & \(4.1024^{+1.1252}_{-0.4414}\) \\
\hline
H1743+322 & \(0.8999^{+0.8169}_{-0.7587}\) & \(11.5716^{+2.7333}_{-2.0382}\) & \(1.0077^{+0.0877}_{-0.1504}\) & \(4.1685^{+0.6888}_{-0.7672}\) \\
\hline
M82 & \(0.5035^{+0.3195}_{-0.3968}\) & \(356.2849^{+2.9560}_{-5.1141}\) & \(0.0172^{+0.0472}_{-0.0153}\) & \(6.0088^{+0.0105}_{-0.0084}\) \\
\hline
Sgr A* & \(1.0071^{+0.8500}_{-0.9231}\) & \((3.61^{+0.12}_{-0.10})\times10^8\) & \(1.0655^{+0.8782}_{-1.0327}\) & \(4.5548^{+3.0112}_{-1.5343}\) \\
\hline
\textbf{Model: WD} &  &  &  &  \\
\hline
XTE & \(0.9635^{+1.2579}_{-0.9400}\) & \(8.3206^{+1.7986}_{-2.0983}\) & \(1.0135^{+0.1178}_{-0.2063}\) & \(5.6849^{+1.2722}_{-0.8592}\) \\
\hline
GRO & \(0.4681^{+1.3979}_{-0.4394}\) & \(6.2654^{+1.5535}_{-1.4859}\) & \(1.1000^{+0.0816}_{-0.2003}\) & \(4.9012^{+1.2304}_{-0.8405}\) \\
\hline
GRS & \(0.9393^{+1.7128}_{-0.8947}\) & \(13.5838^{+3.0245}_{-4.0273}\) & \(0.9646^{+0.2120}_{-0.7734}\) & \(5.7217^{+1.9141}_{-0.9764}\) \\
\hline
H1743+322 & \(1.0418^{+0.9183}_{-1.0061}\) & \(9.5489^{+3.0381}_{-1.8647}\) & \(0.9888^{+0.1612}_{-0.2562}\) & \(5.7628^{+1.0178}_{-1.2502}\) \\
\hline
M82 & \(0.4858^{+0.4979}_{-0.4696}\) & \(365.2850^{+8.6751}_{-11.7964}\) & \(0.0574^{+0.1404}_{-0.0536}\) & \(8.1459^{+0.1321}_{-0.1374}\) \\
\hline
Sgr A* & \(0.9008^{+1.0548}_{-0.7613}\) & \((4.31^{+0.54}_{-0.49}) \times 10^8\) & \(0.5399^{+0.4187}_{-0.5194}\) & \(6.3572^{+3.0840}_{-2.2454}\) \\
\hline
\end{tabular}
\label{tab3}
\end{table*}

Table. \ref{tab3} summarizes the best-fit values and corresponding uncertainties of the NED black hole parameters for other five theoretical model.  The NED parameter \( b \), exhibits slightly varying constraints across all the models studied.  In the PR model, \( b \) takes relatively higher values ranging from approximately \( 0.94 \) to \( 1.88 \), indicating stronger non-linear effects. The ER2 model shows moderately constrained values between \( \sim0.80 \) and \( \sim1.65 \), while ER3 exhibits a noticeable suppression with \( b \) values mostly below 1, suggesting weaker non-linear corrections. The ER4 model offers the tightest constraints, with \( b \in [0.50, 1.09] \), pointing toward a more refined fit. Finally, the WD model demonstrates the broadest variation, where \( b \) spans from \( \sim0.47 \) to \( \sim1.26 \), reflecting diverse source-dependent behavior. Overall, the trend shows a gradual reduction in the NED parameter from PR to ER4, highlighting model dependence and its impact on the effective NED strength.The constraints on the parameter \(b\) exhibit  variations when different models are considered, but these variations remain relatively narrow. Specifically, the value of \(b\) consistently falls within the range of 0 to 2 across all models. \\

\section{Summary and Concluding Remarks}\label{secVI}

\begin{figure*}[t]
    \centering
    \subfloat[GRO J1655-40]{%
        \includegraphics[width=0.45\textwidth]{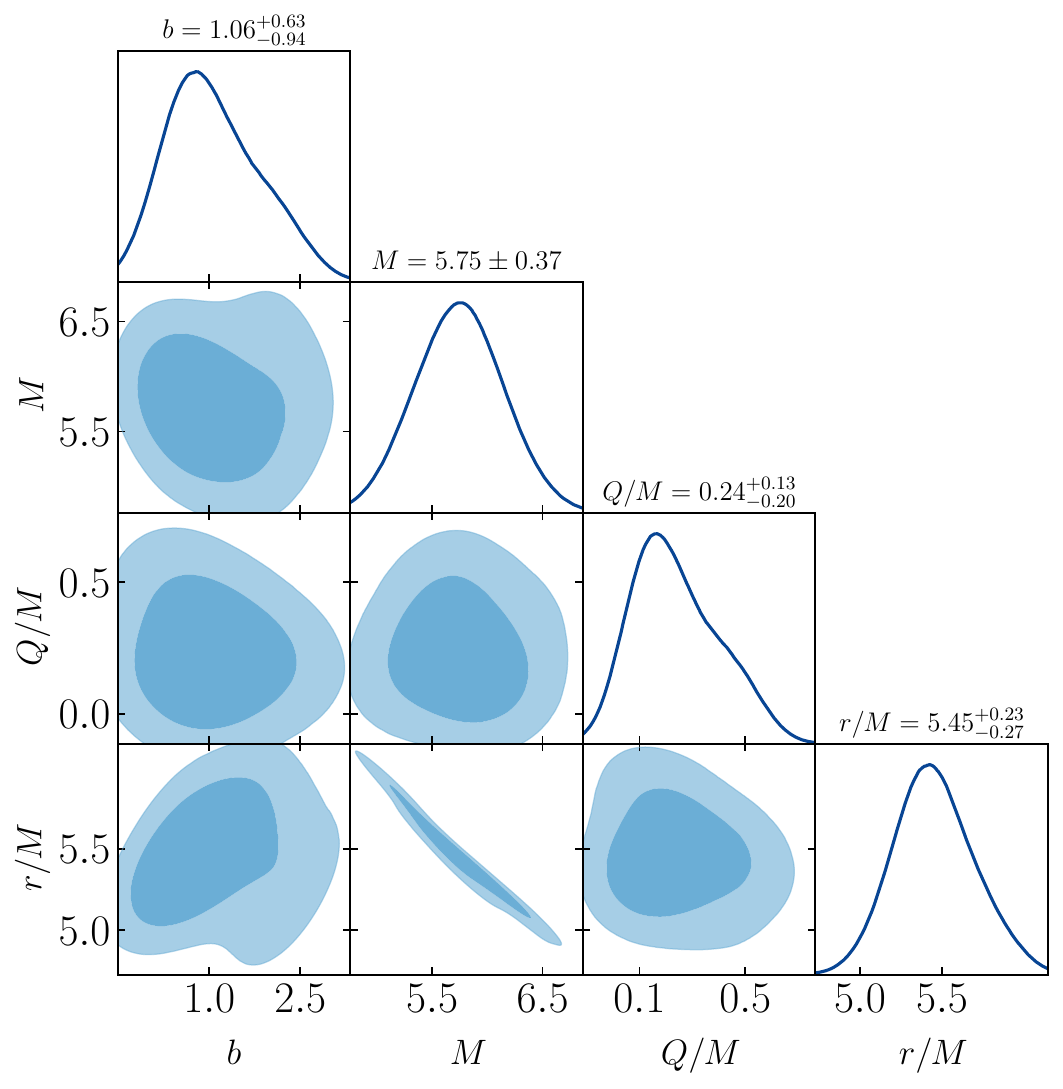}
    }\hfill
    \subfloat[XTE J1550-564]{%
        \includegraphics[width=0.45\textwidth]{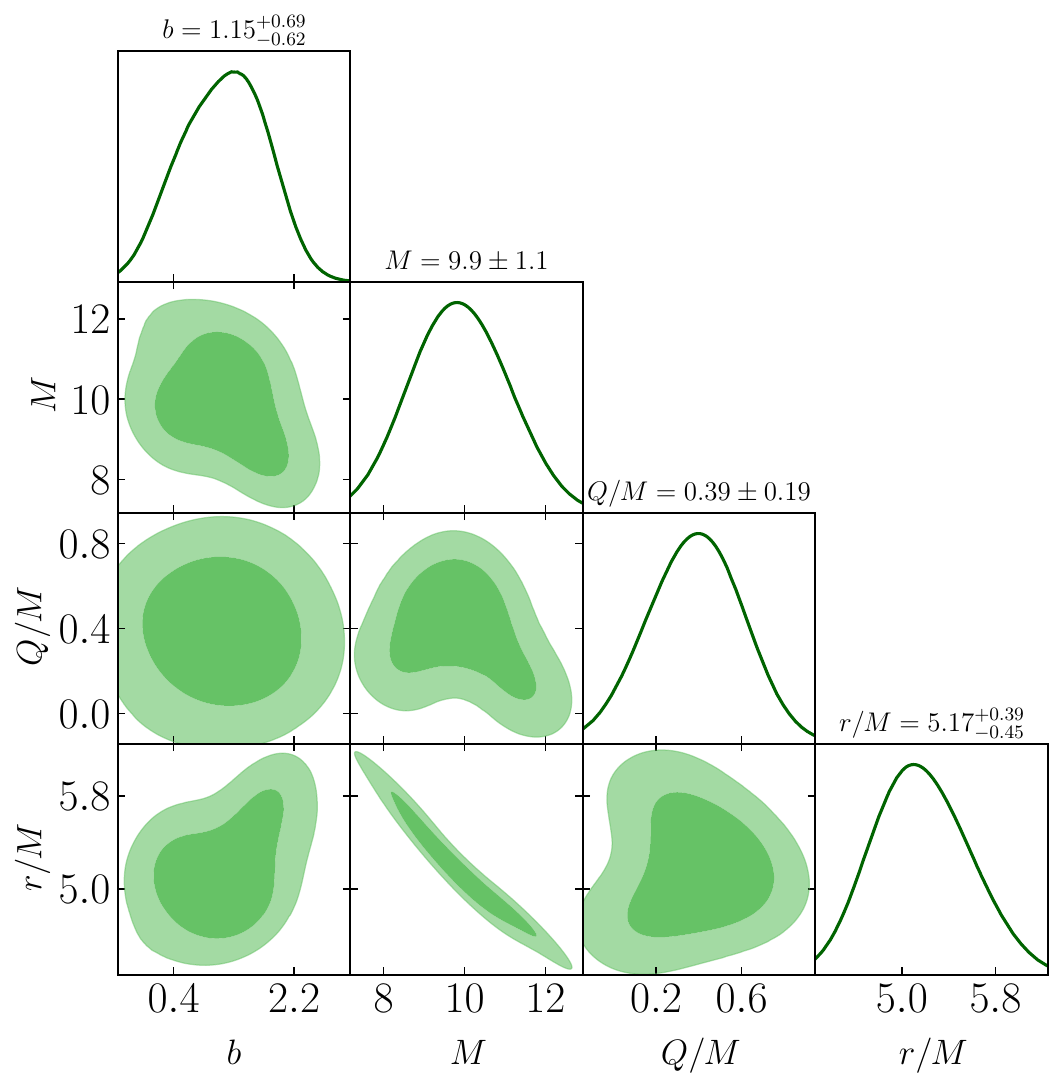}
    }\hfill
    \subfloat[GRS 1915+105]{%
        \includegraphics[width=0.45\textwidth]{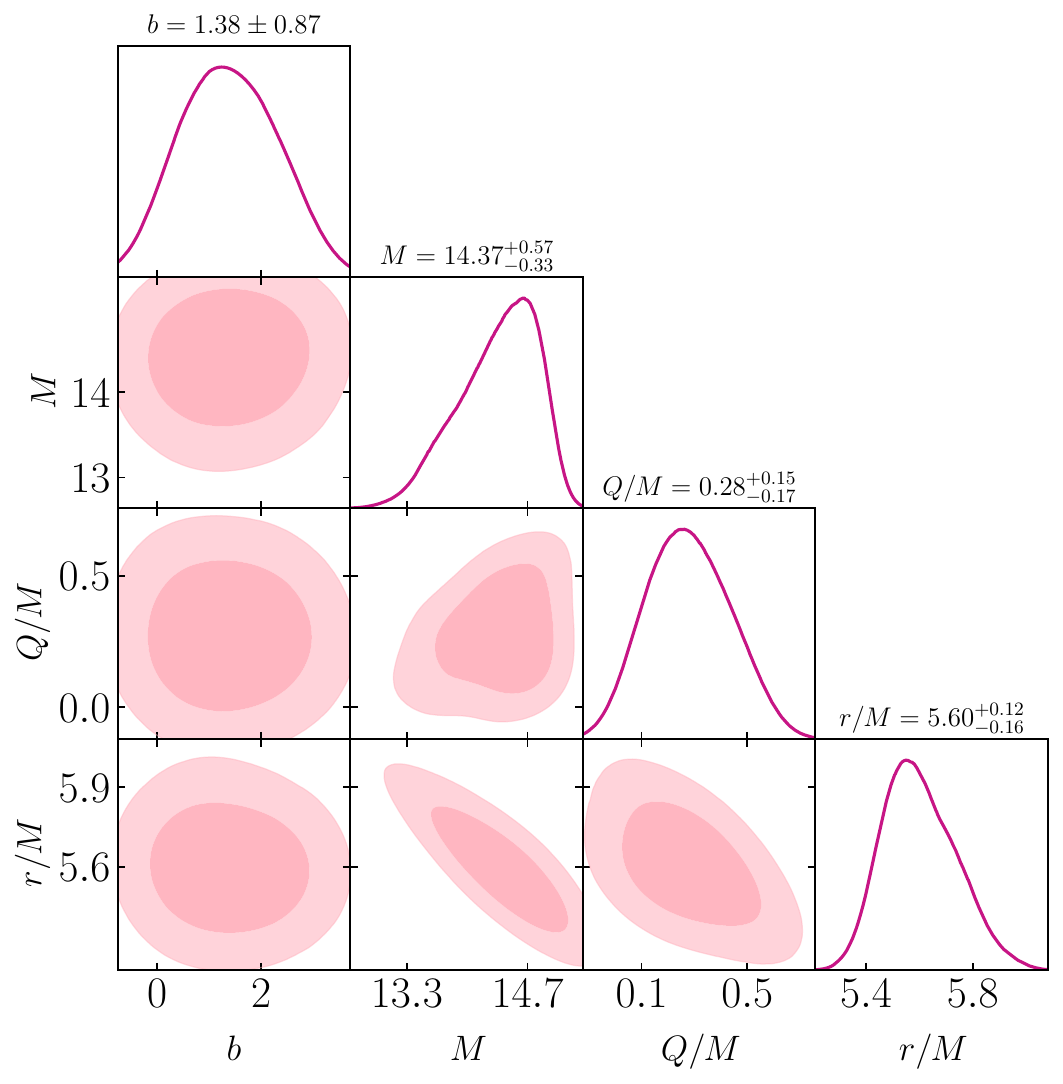}
    }\hfill
    \subfloat[H 1743+322]{%
        \includegraphics[width=0.45\textwidth]{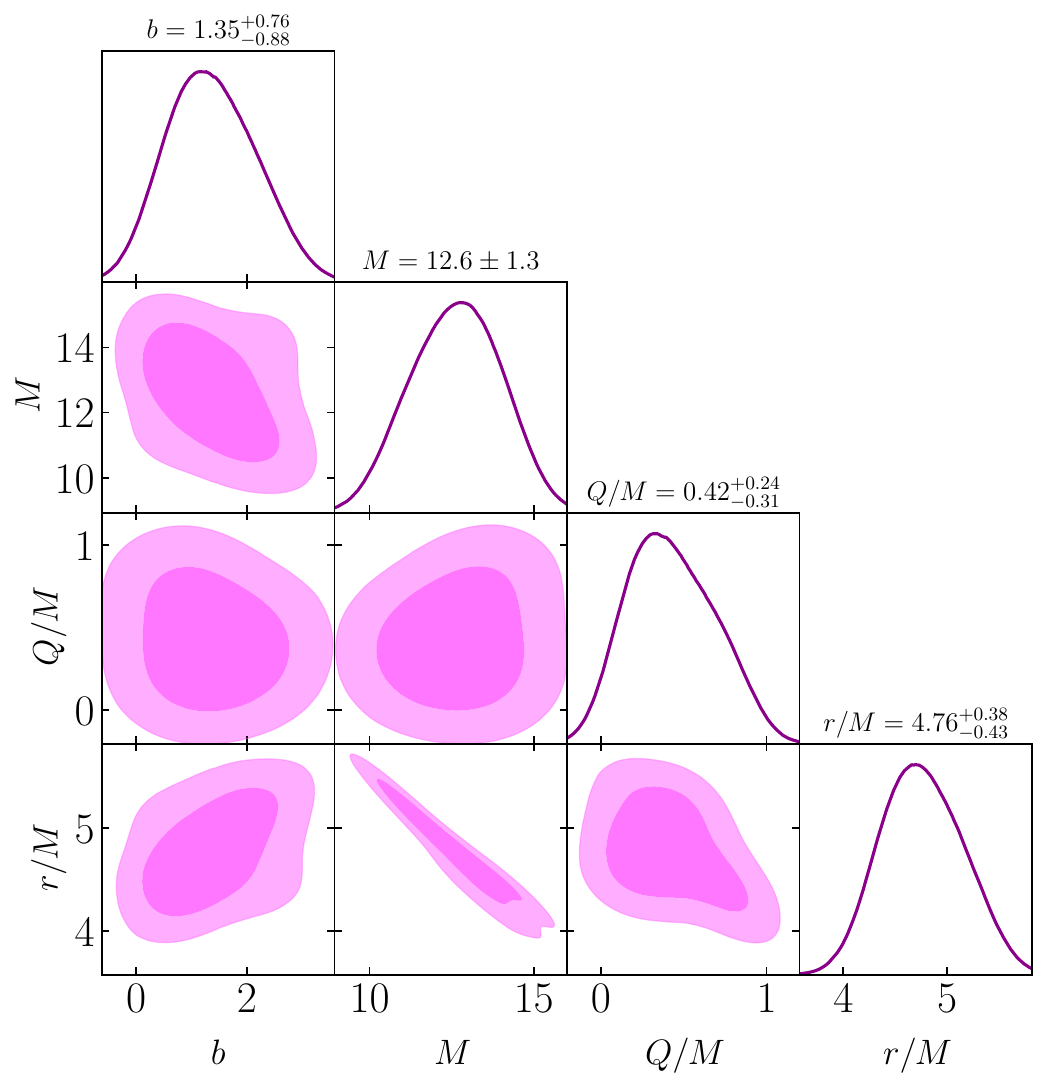}
    }\hfill
    \caption{Corner plots showing posterior distributions for the parameters \(M\), \(b\), \(Q/M\), and \(r/M\) obtained from the MCMC analysis for each black hole source.}
    \label{fig7}
\end{figure*} 

\begin{figure*}[t]
    \centering
    \subfloat[M82 X-1]{%
        \includegraphics[width=0.45\textwidth]{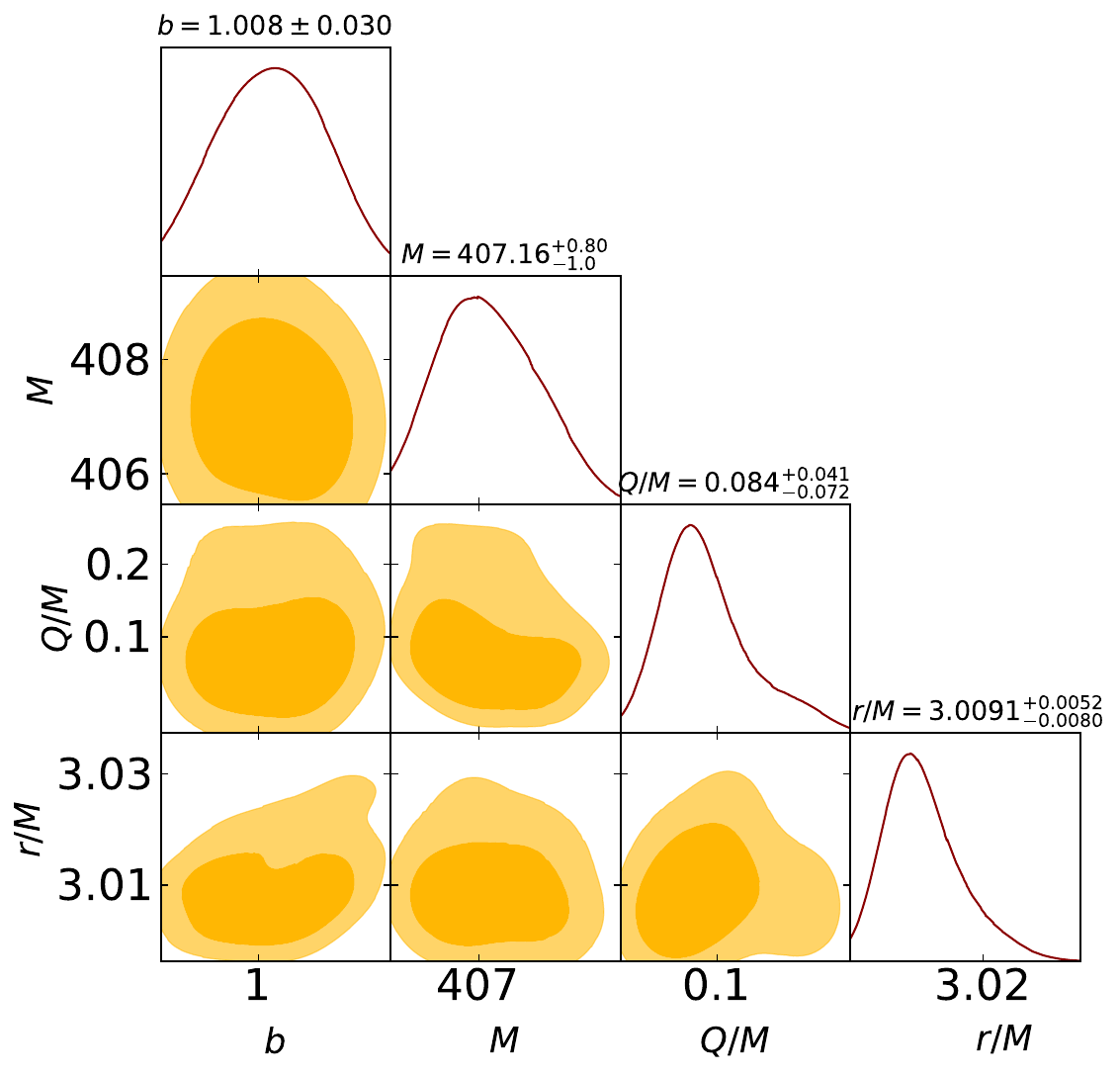}
    }\hfill
    \subfloat[Sgr A*]{%
        \includegraphics[width=0.45\textwidth]{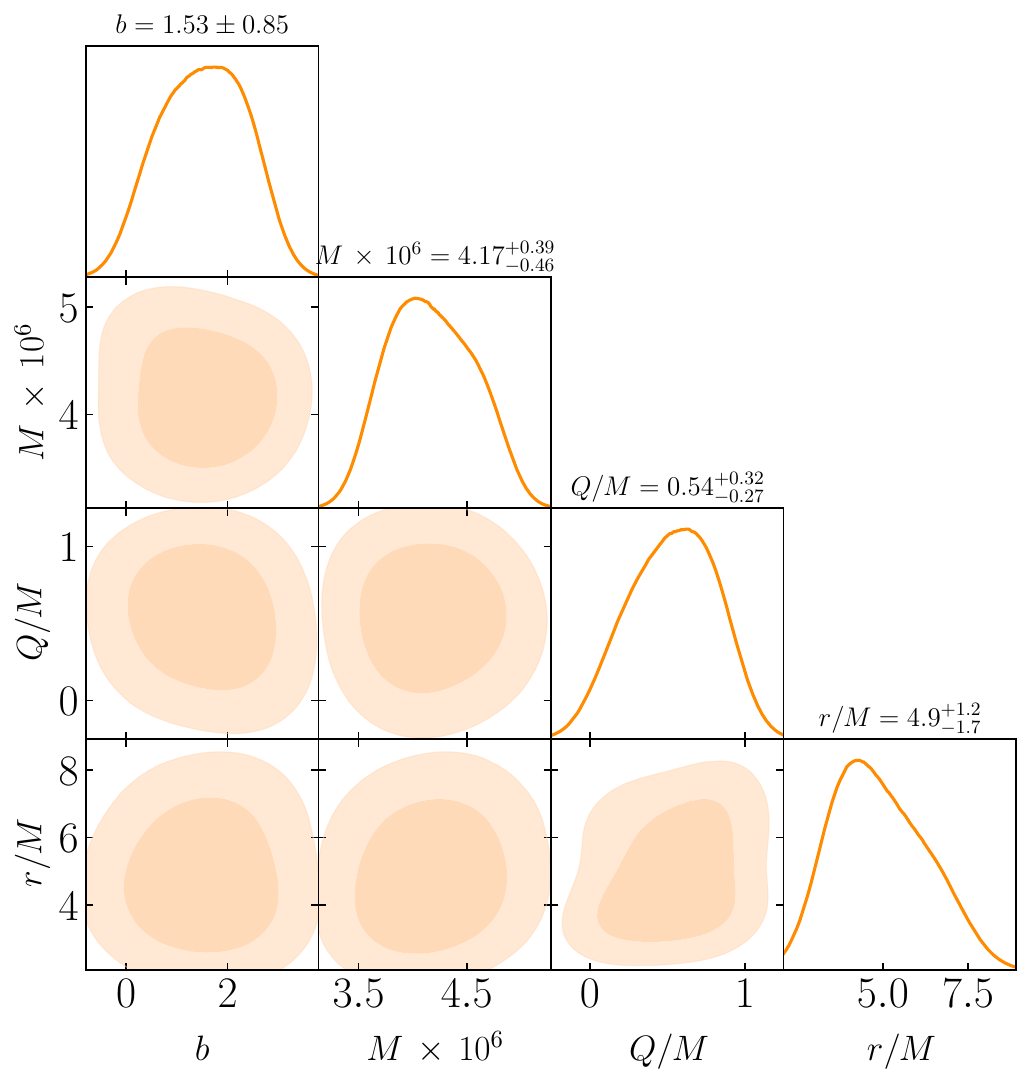}
    }
    \caption{Corner plots showing posterior distributions for the parameters \(M\), \(b\), \(Q/M\), and \(r/M\) obtained from the MCMC analysis for each black hole source.}
    \label{fig7}
\end{figure*}

In this work, we have investigated the circular motion and oscillatory behavior of test particles in the background of a black hole solution influenced by nonlinear electrodynamics (NED), with particular focus on quasi-periodic oscillation (QPO) applications. Starting with the equations of motion, we analyzed the effective potential for circular orbits. Our findings show that, for a neutral test particle, increasing the NED parameter $b$ while keeping the black hole charge $Q$ fixed results in a lower peak of the effective potential compared to the Reissner–Nordström (RN) black hole. As $b$ increases further, the effective potential begins to resemble that of the Schwarzschild case, where charge is absent. 

We also examined the specific energy and angular momentum of particles in circular orbits and found that both quantities decrease with increasing $b$, suggesting that the orbits become more tightly bound. As a consequence, stable circular orbits shift outward. However, a significant drop in these quantities may also reduce the extent of the stable region, making certain orbits more susceptible to instability. The influence of the NED parameter introduces notable deviations from the Schwarzschild and RN cases, highlighting the impact of nonlinear electromagnetic effects on particle dynamics.

Next, we explored the behavior of the innermost stable circular orbit (ISCO) under the influence of NED. For all values of $Q$, the ISCO radius increases monotonically with increasing $b$, indicating that NED effects push the ISCO outward. At fixed $b$, we observed that the ISCO radius increases nearly linearly with the charge parameter $Q$. As expected, in the limit $Q \to 0$, the ISCO approaches the Schwarzschild value of $r_{\text{ISCO}}/M = 6$. Furthermore, we calculated the Keplerian frequency for test particles orbiting the NED black hole and found that it decreases with increasing radial distance $r$. The presence of charge and NED corrections further reduces the orbital frequency $\Omega_\phi$ compared to the Schwarzschild case. Higher values of $b$ lead to a more significant drop in orbital frequency, emphasizing the role of the NED parameter in modifying particle motion. Overall, our analysis shows that as $b$ increases, the dynamical quantities gradually tend toward those found in the Schwarzschild background, illustrating the smooth transition between different black hole geometries.

Furthermore, we extended our study to the investigation of quasi-periodic oscillations (QPOs) by analyzing epicyclic motions within the Parametric resonance(PR), Relativistic Precession (RP), Warped Disk (WD), and Epicyclic Resonance (ER) models. In particular, we focused on resonance conditions corresponding to frequency ratios of 3:2, 4:3, and 5:4 for both upper and lower QPO modes. Our results show that among the examined models, the ER2 configuration produces QPOs at the largest orbital radii, indicating that resonance orbits form farther from the black hole in this case. The WD model follows in this regard, while the ER4 model yields the smallest QPO-generating radii. Notably, for all models considered, the orbital radius at which resonance occurs increases monotonically with the NED parameter $b$, suggesting that greater deviations from standard electrodynamics push the resonance region outward from the black hole horizon.

Finally, we performed a Markov Chain Monte Carlo (MCMC) analysis to constrain $\{M,b,\frac{Q}{M}, \frac{r}{M}\}$ of NED black hole using observational QPO data from six well-known black hole sources. These sources span three distinct mass ranges: stellar-mass black holes (GRO J1655–40, XTE J1550–564, GRS 1915+105, and H 1743–322), the intermediate-mass black hole M82 X-1, and the supermassive black hole Sgr A*. For this analysis, we adopted the RP model with the 3:2 resonance profile as an example to fit the observed QPO frequencies and the results of the analysis for  rest of the five models are tabulated.   A detailed summary of these results are provided in Table~3 (RP) and Table~3 (Other five models).\\

As our final remark, we emphasize that the NED black hole solution studied in this work behaves as expected in various limiting cases. Specifically, in the absence of charge, the solution smoothly reduces to the Schwarzschild metric, while in the limit \( b \rightarrow 0 \), it recovers the standard RN  solution. Interestingly, in the regime of very large \( b \), the spacetime also asymptotically tends toward the Schwarzschild  profile, indicating a transitional nature of the parameter \( b \) in interpolating between these two classical black hole This theoretical expectation is consistently reflected in our analysis: beginning with the behavior of the effective potential, and continuing through the energy and angular momentum profiles, we observe that increasing \( b \) gradually shifts the dynamics closer to the Schwarzschild characteristics. Conversely, for small \( b \), the black hole mimics RN-like features more prominently. This trend persists in our study of the Keplerian frequency and the innermost stable circular orbit (ISCO), where larger values of \( b \) push the ISCO radius and orbital frequency closer to the Schwarzschild benchmark (\( r_{\text{ISCO}} = 6M \)). More pronounced deviations appear in our investigation of QPO-generating orbits. Across all QPO models considered (RP, WD, and ER), the resonance radius increases monotonically with \( b \), signaling that nonlinear electrodynamics effects tend to shift these orbits outward from the black hole. As \( b \) grows, the QPO profiles steadily approach the Schwarzschild behavior, while for small \( b \), they align well with the RN characteristics.To further solidify these observations, we employed a Markov Chain Monte Carlo (MCMC) analysis to constrain the model parameters using QPO data from a diverse set of black hole sources. Notably, the inferred bounds on the NED parameter \( b \) remained consistently of order unity across all mass regimes—stellar, intermediate, and supermassive. This narrow and stable constraint suggests that nonlinear corrections inherent to NED may indeed influence the observed QPO frequencies in a measurable way.\\

In conclusion, our study clearly demonstrates that the QPO characteristics of the NED black hole interpolate between the RN and Schwarzschild profiles, with the parameter \( b \) playing a pivotal role in governing this transition. Notably, both extremal limits, \( b \rightarrow 0 \) and \( b \rightarrow \infty \), produce well-behaved and physically meaningful profiles,. The most important outcome of our analysis is the clear signature of nonlinear electrodynamics (NED) on the QPO behavior of charged black holes. 
This signature is reflected both in the theoretical aspects such as the effective potential, ISCO, and Keplerian frequency and in the observational context through the MCMC analysis based on QPO data. The constraints obtained on the NED parameter \( b \) remain consistently of order unity across different black hole sources, suggesting that the NED-induced modifications leave a subtle yet discernible signature on the QPO characteristics. These findings may indicate  the relevance of NED corrections in the context of black hole. \\

It is important to mention that although the observational data employed in our analysis pertain to rotating black holes, we have deliberately neglected the spin parameter in our modeling in order to isolate the pure influence of nonlinear electrodynamics (NED) on the quasi-periodic oscillation (QPO) frequencies. Our goal was to investigate whether the introduction of a NED parameter can effectively reproduce phenomenological features commonly attributed to spin, thereby offering an alternative mechanism for explaining the observed QPO behavior. In this context, we speculate that a moderate NED charge may compensate for the frequency shifts typically induced by rotation, allowing black holes without spin to mimic the observational signatures of their rotating counterparts. 
However, the constraints on the NED parameter \(b\) are expected to be influenced by the inclusion of the spin parameter in our framework, as demonstrated in Ref.\cite{c2}, where the interplay between the NED parameter \(k\) and the spin parameter \(a\) in modeling QPOs is examined. The study highlights how the constraints on each parameter are sensitive to the value of the other, indicating a coupled dependence between spin and nonlinear electrodynamics effects.  Notably, incorporating rotation in our case would require extending the background metric to a magnetically charged, rotating black hole solution within the framework of NED which is both technically nontrivial. We intend to explore this in future studies.
\section*{Acknowledgements} 
BH would like to thank DST-INSPIRE, Ministry of Science and Technology fellowship program, Govt. of India for awarding the DST/INSPIRE Fellowship[IF220255] for financial support.  We would also like to extend our thanks to the anonymous referee for the helpful comments and suggestions.

\begin{widetext}
\section{Appendix}\label{A}
\begin{multline}\nonumber
\Omega_r^2=\frac{1}{4 r^6 b^6}\bigl(-2 b M + Q^2 \tanh\bigl(\tfrac{b}{r}\bigr) + b r\bigr)^3 
\biggl(
\frac{
3 b \bigl(2 b M r - Q^2 r \tanh\bigl(\tfrac{b}{r}\bigr) - b Q^2 \text{sech}^2\bigl(\tfrac{b}{r}\bigr)\bigr)
}{
r^2 \bigl(-2 b M + Q^2 \tanh\bigl(\tfrac{b}{r}\bigr) + b r\bigr)^2
}
\\- b \biggl[
r \biggl(
\frac{
2 \bigl(b - \tfrac{b Q^2 \text{sech}^2\bigl(\tfrac{b}{r}\bigr)}{r^2}\bigr)^2
}{
\bigl(-2 b M + Q^2 \tanh\bigl(\tfrac{b}{r}\bigr) + b r\bigr)^3
}
- 
\frac{
\tfrac{2 b Q^2 \text{sech}^2\bigl(\tfrac{b}{r}\bigr)}{r^3}
- \tfrac{2 b^2 Q^2 \tanh\bigl(\tfrac{b}{r}\bigr) \text{sech}^2\bigl(\tfrac{b}{r}\bigr)}{r^4}
}{
\bigl(-2 b M + Q^2 \tanh\bigl(\tfrac{b}{r}\bigr) + b r\bigr)^2
}
\biggr)
- 
\frac{
2 \bigl(b - \tfrac{b Q^2 \text{sech}^2\bigl(\tfrac{b}{r}\bigr)}{r^2}\bigr)
}{
\bigl(-2 b M + Q^2 \tanh\bigl(\tfrac{b}{r}\bigr) + b r\bigr)^2
}
\biggr]
\biggr)
\end{multline}
\end{widetext}

\end{document}